\documentclass[a4paper,fleqn,usenatbib]{mnras}

\usepackage{newtxtext,newtxmath}

\usepackage[T1]{fontenc}
\usepackage{ae,aecompl}

\usepackage{graphicx}	
\usepackage{amsmath}	
\usepackage{amssymb}	

\newcommand{\NICER}{{\it NICER}}
\newcommand{\RXTE}{{\it RXTE}}
\newcommand{\Chandra}{{\it Chandra}}
\newcommand{\XMM}{{\it XMM-Newton}}
\newcommand{\psr}{PSR~J0537$-$6910}
\newcommand{\nus}{\nu}
\newcommand{\nudot}{\dot{\nus}}
\newcommand{\nuddot}{\ddot{\nus}}
\newcommand{\nig}{n_{\rm ig}}
\newcommand{\tig}{\tau_{\rm ig}}
\newcommand{\tobs}{t_{\rm obs}}
\newcommand{\Ag}{A_{\rm g}}
\newcommand{\tauc}{\tau_{\rm c}}
\newcommand{\Isf}{I_{\rm sf}}
\newcommand{\mn}{m_{\rm n}}
\newcommand{\mneff}{\left<\mn^\ast\right>}
\newcommand{\nugw}{\nu_{\rm gw}}
\newcommand{\hsd}{h_{\rm sd}}

\title[\NICER\ observations of \psr]{Return of the Big Glitcher: \NICER\ timing and glitches of \psr}

\author[W. C. G. Ho et al.]{Wynn C. G. Ho$^{1}$\thanks{E-mail: wynnho@slac.stanford.edu},
Crist\'obal M. Espinoza$^{2}$,
Zaven Arzoumanian$^{3}$,
Teruaki Enoto$^{4}$,
\newauthor
Tsubasa Tamba$^{5}$,
Danai Antonopoulou$^{6}$,
Micha{\l} Bejger$^{6}$,
Sebastien Guillot$^{7,8}$,
\newauthor
Brynmor Haskell$^{6}$,
Paul S. Ray$^{9}$
\\
$^{1}$Department of Physics and Astronomy, Haverford College, 370 Lancaster Avenue, Haverford, PA, 19041, USA\\
$^{2}$Departamento de F\'isica, Universidad de Santiago de Chile, Avenida Ecuador 3493, 9170124 Estaci\'on Central, Santiago, Chile\\
$^{3}$X-Ray Astrophysics Laboratory, NASA Goddard Space Flight Center, Greenbelt, MD, 20771, USA\\
$^{4}$Extreme Natural Phenomena RIKEN Hakubi Research Team, RIKEN Cluster for Pioneering Research, 2-1 Hirasawa, Wako, Saitama, 351-0198, Japan\\
$^{5}$Department of Physics, University of Tokyo, 7-3-1 Hongo, Bunkyo-ku, Tokyo, 113-0033, Japan\\
$^{6}$Nicolaus Copernicus Astronomical Center, Polish Academy of Sciences, ul. Bartycka 18, 00-716 Warsaw, Poland\\
$^{7}$IRAP, CNRS, 9 avenue du Colonel Roche, BP 44346, F-31028 Toulouse Cedex 4, France\\
$^{8}$Universit\'e de Toulouse, CNES, UPS-OMP, F-31028 Toulouse, France\\
$^{9}$Space Science Division, U.S. Naval Research Laboratory, Washington, DC, 20735, USA\\
}

\date{Accepted 2020 August 26. Received 2020 August 26; in original form 2020 June 18}

\pubyear{2020}

\begin{document}
\label{firstpage}
\pagerange{\pageref{firstpage}--\pageref{lastpage}}
\maketitle

\begin{abstract}
\psr, also known as the Big Glitcher, is the most prolific glitching
pulsar known, and its spin-induced pulsations are only detectable in X-ray.
We present results from analysis of 2.7~years of \NICER\ timing
observations, from 2017 August to 2020 April.
We obtain a rotation phase-connected timing model for the entire
timespan, which overlaps with the third observing run of LIGO/Virgo,
thus enabling the most sensitive gravitational wave
searches of this potentially strong gravitational wave-emitting pulsar.
We find that the short-term braking index between glitches decreases
towards a value of 7 or lower at longer times since the preceding glitch.
By combining \NICER\ and \RXTE\ data,
we measure a long-term braking index $n=-1.25\pm0.01$.
Our analysis reveals 8 new glitches, the first detected since 2011,
near the end of \RXTE, with a total \NICER\ and \RXTE\ glitch
activity of $8.88\times 10^{-7}\mbox{ yr$^{-1}$}$.
The new glitches follow the seemingly unique time-to-next-glitch---glitch-size
correlation established previously using \RXTE\ data,
with a slope of $5\mbox{ d $\mu$Hz$^{-1}$}$.
For one glitch around which \NICER\ observes two days on either side,
we search for but do not see clear evidence of spectral nor pulse
profile changes that may be associated with the glitch.
\end{abstract}




\section{Introduction} \label{sec:intro}

With a spin frequency $\nus\approx62\mbox{ Hz}$
(spin period $P\approx 16\mbox{ ms}$),
\psr\ is the fastest-rotating young pulsar known
and is located in the 1--5~kyr old supernova remnant N157B
\citep{wangetal98,chenetal06} in the Large Magellanic Cloud at a
distance of 49.6~kpc \citep{pietrzynskietal19}.
Its spin rate is only measurable at X-ray and gamma-ray energies up
to $\sim$60~keV \citep{marshalletal98,kuiperhermsen15}, and the
pulsar has the highest spin-down energy loss rate
$\dot{E}=4.9\times 10^{38}\mbox{ erg s$^{-1}$}$ among more than
2800 known pulsars \citep{manchesteretal05}.
While the pulsar's spin frequency decreases over the entire
13 years of \RXTE\ observation from 1999--2011
at a rate $\nudot\approx-1.99\times 10^{-10}\mbox{ Hz s$^{-1}$}$,
\psr\ underwent a remarkable 42 or 45 spin-up glitches,
yielding an average glitch rate of $>3.2\mbox{ yr$^{-1}$}$
\citep{marshalletal04,middleditchetal06,antonopoulouetal18,ferdmanetal18},
and its glitch sizes are larger than those seen in most glitching
pulsars \citep{espinozaetal11,yuetal13,fuentesetal17,hoetal20}.
Because of its high glitch activity, \psr\ proves to be extremely
useful in theoretical understanding of the mechanism that produces glitches
\citep{linketal99,melatosetal08,anderssonetal12,chamel13,hoetal15},
which is thought to be due to unpinning of superfluid vortices in the
star's crust and possibly its core \citep{andersonitoh75,alparetal84}.

What makes \psr\ even more extraordinary is the predictability of
when its glitches occur.
\citet{middleditchetal06,antonopoulouetal18,ferdmanetal18} measure
a linear correlation between glitch size $\Delta\nus$ and time to
next glitch, with a slope of $\sim 0.2\mbox{ $\mu$Hz d$^{-1}$}$
and prediction accuracy of days (see Section~\ref{sec:predict}).
Such a clear correlation does not seem to hold for any other pulsar
and may be unique to \psr\ \citep{melatosetal18}.
\citet{fuentesetal19} find the large glitches
of the Vela pulsar show a weak correlation,
\citet{akbaletal17} predict Vela glitch times to an accuracy of about
$\pm0.4$~yr,
and \citet{melatosdrummond19} predict the next glitch for three other
pulsars but with large uncertainties of $\pm0.7$, 4, and 5~yr.
Furthermore, statistical analyses give some indication that
glitch sizes and times to next glitch each show a bimodal
distribution in the case of \psr\ \citep{howittetal18}.
Glitch times also appear quasi-periodic \citep{middleditchetal06,melatosetal08}.

\psr\ is of additional interest because of its potential as a
source of detectable gravitational waves (GWs).
The frequencies at which \psr\ might emit GWs are in the most
sensitive frequency band of ground-based detectors
(around 100~Hz; \citealt{abbottetal19c}).
There is also tantalizing but speculative evidence shown by
\citet{anderssonetal18} for the generation of GWs in this pulsar
by a stellar oscillation, i.e., r-mode oscillation
\citep{andersson98,friedmanmorsink98,anderssonkokkotas01}.
The most sensitive searches for continuous GW emission from known pulsars
use contemporaneous electromagnetic observations to track a pulsar's
spin evolution and thereby reduce the large parameter
space of a search \citep{abbottetal19c}.
Such a targeted search has not been done in the advanced detector era
for \psr\ (cf. narrow-band search; \citealt{fesikpapa20,fesikpapa20b})
because of the lack of a timing model to compare to GW data,
and an accurate phase-connected model is not possible over long times
without monitoring because of the pulsar's high glitch rate.

With the demise of \RXTE\ in 2012, the 42 glitches of \psr\ measured by
\citet{ferdmanetal18} or 45 glitches measured by \citet{antonopoulouetal18}
are all the ones obtainable from existing data prior to \NICER;
the difference in glitch numbers is due to small glitches just
above or below detection thresholds.
\NICER\ began observing \psr\ soon after launch in 2017 June.
As we show here, \NICER\ clearly detects the pulsar's spin rate, with
typical pulse time-of-arrival (TOA) uncertainties ($<100\mbox{ $\mu$s}$)
generally better than those obtained using \RXTE.
In Section~\ref{sec:data}, we describe \NICER\ observations of \psr\
and timing analysis of these observations.
In Section~\ref{sec:spinevol}, we present our timing model and
measurements of braking indices of \psr.
In Section~\ref{sec:glitch0}, we discuss measured glitches and their
properties.
In Section~\ref{sec:discuss}, we summarize and discuss some
implications of our results.

\section{\NICER\ data} \label{sec:data}

We process and filter \NICER\ data on \psr\
using HEASoft~6.22--6.26 and NICERDAS~2018-03-01\_V003--2020-01-08\_V006c.
We exclude all events from ``hot'' detector 34, which gives elevated
count rates in some circumstances, and portions of exposure accumulated
during passages through the South Atlantic Anomaly.
While \NICER\ is sensitive to 0.25--12~keV photons,
we make an energy cut and extract only events from 1--7~keV,
where pulsations are easily detected
(see, e.g., \citealt{marshalletal98,kuiperhermsen15}).
We ignore time intervals of enhanced background affecting all
detectors by constructing a light curve binned at 16~s and
removing intervals strongly contaminated by background flaring
when the count rate exceeds $10\mbox{ c s$^{-1}$}$.
\NICER\ experienced a time stamp anomaly which resulted in incorrect
time stamps for data taken with MPU1 between 2019 July 8 and 23;
we follow the recommended procedure for excluding MPU1 data
within this time window\footnote{\url{https://heasarc.gsfc.nasa.gov/docs/nicer/data_analysis/nicer_analysis_tips.html\#July2019-MPU1_Timing_Errors}}.
Using these filtering criteria, we obtain clean data with
count rates $\sim 3\mbox{ c s$^{-1}$}$ for use in pulse timing analysis.
We do not conduct spectral analyses using \NICER\ data, except for
a small subset (see Section~\ref{sec:variability}), since the
large non-imaging field of view implies an extracted spectrum that
will primarily be due to that of the supernova remnant and such spectra
from X-ray imaging telescopes are presented in other studies
\citep{chenetal06,kuiperhermsen15}.

We combine sets of individual ObsIDs into merged observations, with
each merged observation yielding a single time-of-arrival (TOA) measurement.
ObsIDs are combined such that there is sufficient exposure to
confidently detect the spin frequency of \psr, with typical total
exposures of 4--9~ks (see below), and merged ObsIDs are those acquired
usually within a 3--4~day span and on rare occasions within 6--7~days.
We obtain 95 merged observations for the data and timespan
(2017 August 17 to 2020 April 25) presented here\footnote{The first
\NICER\ observations of \psr\ with significant exposure time occur
in 2017 July.  While we are able to construct 3 TOAs from these
observations, we are unable to obtain a sensible phase-connection
between these TOAs and others that follow them.  This may be due to a
small magnitude glitch (with $\Delta\nus<0.06\mbox{ $\mu$Hz}$)
between MJDs 57963 and 57984, but we are unable to
determine this conclusively or characterize this possible glitch
because of the insufficient number of TOAs provided by the 2017 July data.}.
Before performing a pulsation search, we use \texttt{barycorr} to
transform between Terrestrial Time, used for event time stamps, and
Barycentric Dynamical Time (TDB).
We adopt the JPL~DE421 solar system ephemeris and the \Chandra\
sky position of \psr\ measured by \citet{townsleyetal06}, i.e.,
R.A.$=05^{\rm h}37^{\rm m}47.\!\!^{\rm s}416$,
decl.$=-69^\circ10\arcmin19.\!\!\arcsec88$ (J2000);
note the $0.\!\!\arcsec4\pm0.\!\!\arcsec2$ (1$\sigma$) difference
in position from that measured by \citet{chenetal06}.

\begin{figure}
\begin{center}
\includegraphics[width=\columnwidth]{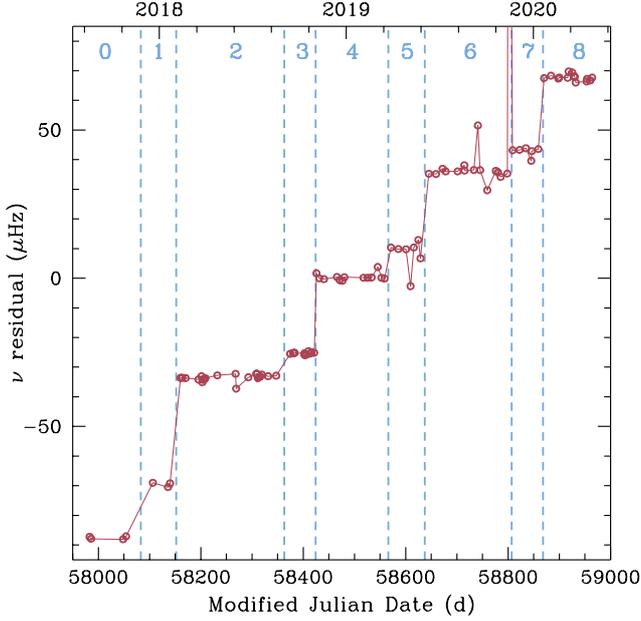}
\caption{
Difference between the measured candidate spin frequency of \psr\
and linear model for frequency evolution, $\nus_4+\nudot_4(t-t_4)$,
where $\nu_4$ and $\nudot_4$ are values from segment 4 at
$t_4=\mbox{MJD }58493$ (see Table~\ref{tab:data}).
Segments are labeled by numbers and separated by the occurrence
of a glitch, each of which is denoted by a vertical dashed line.
}
\label{fig:f0resid}
\end{center}
\end{figure}

\begin{figure}
\begin{center}
\includegraphics[width=\columnwidth]{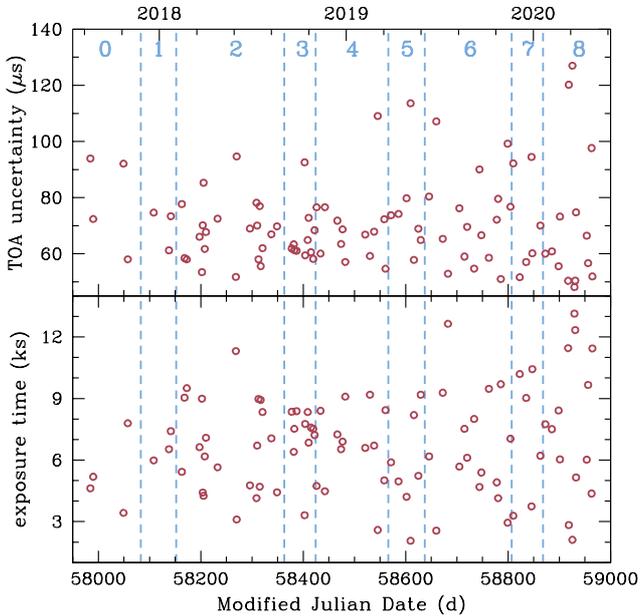}
\caption{
Uncertainty in measured time-of-arrival (top) and total exposure
time (bottom) for each merged observation.
Segments are labeled by numbers and separated by the occurrence
of a glitch, each of which is denoted by a vertical dashed line.
}
\label{fig:toa}
\end{center}
\end{figure}

Acceleration searches are conducted using PRESTO \citep{ransometal02},
with searches using a time bin of 0.5~ms and usually including
8 harmonics given the narrow pulse profile of \psr\
(see, e.g., Section~\ref{sec:variability}).
Pulsations at the spin frequency ($\nus\approx 61.9\mbox{ Hz}$)
are almost always the strongest detected.
Data are folded at the candidate pulse frequency using \texttt{prepfold}
and a refined frequency is determined.
On occasion, further iterations are performed to obtain a more
robust measurement.
Figure~\ref{fig:f0resid} shows the difference between the candidate spin
frequency measurement from each merged observation and a simple linear model
of frequency evolution, $\nus_4+\nudot_4(t-t_4)$, arbitrarily set to values
obtained from the fourth segment of data.
The slope evident in some segments is due to a difference in actual
$\nudot$ for that segment.
Each segment is separated by the occurrence of a glitch.
Finally, we determine the TOA for each merged observation by
cross-correlation with a template pulse profile generated from fitting
a Gaussian to a series of \NICER\ pulse profiles of \psr.
Figure~\ref{fig:toa} shows the measured uncertainty of each TOA
(top panel) and total exposure time of each merged observation
that is used for each TOA measurement (bottom panel).

\section{Spin evolution} \label{sec:spinevol}
\subsection{Timing model} \label{sec:timing}

\begin{figure}
\begin{center}
\includegraphics[width=\columnwidth]{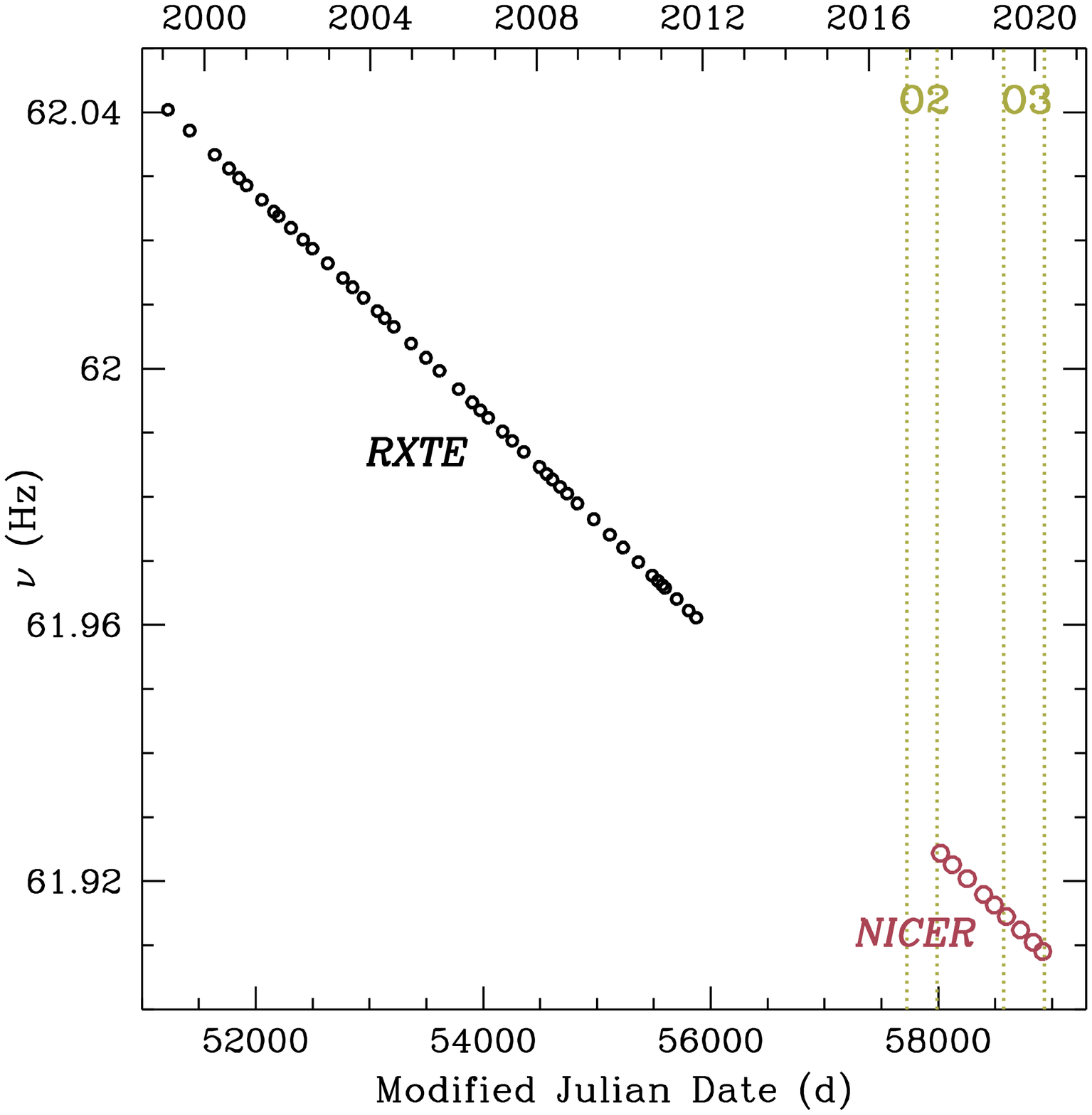}
\caption{
Evolution of the spin frequency $\nus$ of \psr.
Large circles are \NICER\ values measured by fitting a timing model
to TOAs in each segment (see Table~\ref{tab:data}).
Small circles denote \RXTE\ values from Table~1 of \citet{antonopoulouetal18}.
Vertical dotted lines denote the
start (MJD 57722, 2016 November 30) and
end (MJD 57990, 2017 August 25) of the second observing run (O2) and
start (MJD 58574, 2019 April 1) and
end (MJD 58935, 2020 March 27) of the third observing run (O3) of LIGO/Virgo.
}
\label{fig:tf0}
\end{center}
\end{figure}

\begin{figure}
\begin{center}
\includegraphics[width=\columnwidth]{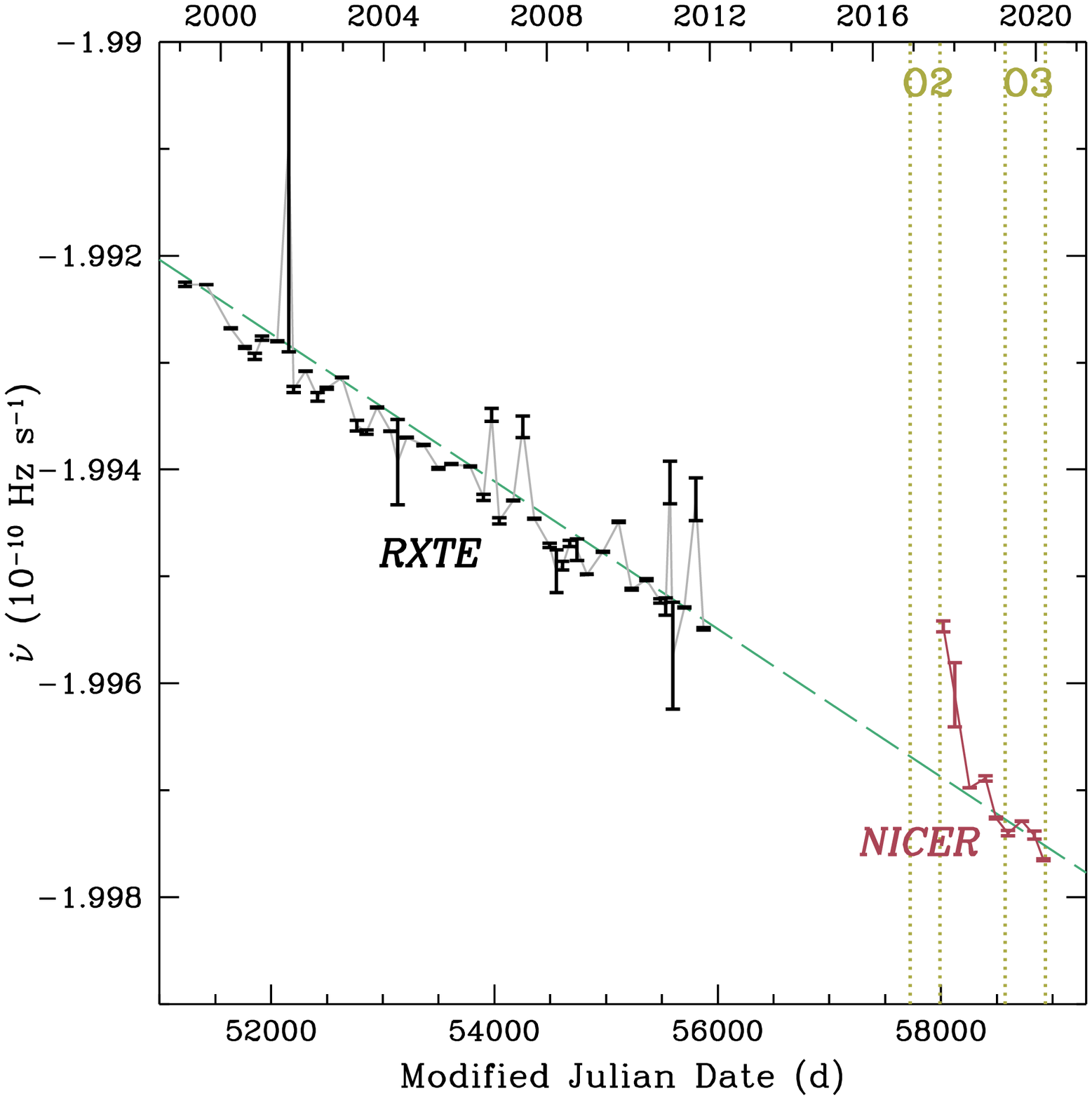}
\caption{
Evolution of the spin frequency time derivative $\nudot$ of \psr.
\NICER\ values are measured by fitting a timing model
to TOAs in each segment (see Table~\ref{tab:data}).
\RXTE\ values are from Table~1 of \citet{antonopoulouetal18}.
Errors are 1$\sigma$ uncertainty.
Dashed line shows a linear fit of \NICER\ and \RXTE\ data with
best-fit $\nuddot=-8.00\times 10^{-22}\mbox{ Hz s$^{-2}$}$.
Vertical dotted lines denote the
start (MJD 57722, 2016 November 30) and
end (MJD 57990, 2017 August 25) of the second observing run (O2) and
start (MJD 58574, 2019 April 1) and
end (MJD 58935, 2020 March 27) of the third observing run (O3) of LIGO/Virgo.
}
\label{fig:tf1}
\end{center}
\end{figure}

We use TEMPO2 \citep{hobbsetal06} to fit TOAs in each segment between
glitches with a timing model that includes a fiducial phase and the
pulsar spin frequency $\nu$ and its first and second time derivatives,
$\nudot$ and $\nuddot$, as free parameters.
However, for segments 0, 1, and 5, we fix $\nuddot$ at
$10^{-20}\mbox{ Hz s$^{-2}$}$
[i.e., mean value during segments between glitches in the \RXTE\ era
\citep{antonopoulouetal18} and is generally consistent with our
\NICER\ measurements]
because there are insufficient numbers of TOAs to fully constrain the model.
Furthermore, for segment 1 when we only have 3 TOAs, $\nus$ and $\nudot$ are
determined by searching within a range of values and selecting
those that yield the lowest root-mean-square (RMS) residuals.
Specifically, the search is performed centred at the first candidate
$\nus$ measurement in the segment (see Figure~\ref{fig:f0resid}) and
at the $\nudot$ expected from the general long-term trend
(see Figure~\ref{fig:tf1}), with search radii of $2\mbox{ $\mu$Hz}$
and $1\times10^{-13}\mbox{ Hz s$^{-1}$}$.
Quoted uncertainties correspond to half the range of values covered
by all solutions that yield a RMS lower than $75\mbox{ $\mu$s}$,
which is the largest measured TOA uncertainty in segment~1.
The results of these timing model fits are given in Table~\ref{tab:data},
with quoted uncertainties being the formal 1$\sigma$ errors from the fits.
Figures~\ref{fig:tf0} and \ref{fig:tf1} show the spin frequency
$\nus$ and its time derivative $\nudot$ for each segment,
as well as interglitch values measured using \RXTE.

\begin{table*}
\centering
\caption{\psr\ timing model parameters for segments between glitch epochs.
Number in parentheses is 1$\sigma$ uncertainty in last digit.}
\label{tab:data}
\begin{tabular}{cccccllcccc}
\hline
Segment & Epoch & Start & End & TOAs & $\qquad\quad\nus$ & $\quad\qquad\nudot$ & $\nuddot$ & $\nig$ & Residual RMS & $\chi^2/\mbox{dof}$ \\
 & (MJD) & (MJD) & (MJD) & & $\,\qquad$(Hz) & ($10^{-10}$ Hz s$^{-1}$) & ($10^{-20}$ Hz s$^{-2}$) & & ($\mu$s) & \\
\hline
0 & 58020 & 57984.5 & 58057.5 &  4 & 61.924374260(2)   & $-1.99547(5)$  & $[1]^a$   &     --- & 66.14 &  3.1 \\
1 & 58124 & 58108.1 & 58141.6 &  3 & 61.922597203(3) & $-1.9961(3)$  & $[1]^a$   &     --- & 1.158 &  ---$^b$ \\
2 & 58255 & 58163.0 & 58348.8 & 21 & 61.9203729962(9)  & $-1.996977(2)$ & 0.56(1) &  8.7(2) & 123.5 &  4.4 \\
3 & 58399 & 58377.5 & 58422.2 & 11 & 61.917896251(3)   & $-1.99689(2)$  & 5.9(6)  &  90(10) & 56.39 &  1.2 \\
4 & 58493 & 58426.1 & 58560.6 & 13 & 61.916299559(4)   & $-1.997261(8)$ & 0.81(8) &   13(1) & 250.8 &  20 \\
5 & 58600 & 58571.3 & 58629.3 &  7 & 61.914462421(2)   & $-1.99740(3)$  & $[1]^a$   &     --- & 98.49 &  3.3 \\
6 & 58723 & 58645.5 & 58804.5 & 16 & 61.912366620(2)   & $-1.997289(4)$ & 0.88(3) & 13.7(4) & 142.7 &  5.9 \\
7 & 58836 & 58810.0 & 58862.9 &  6 & 61.910424210(6)   & $-1.99742(4)$  & 5.8(9)  &  90(10) & 81.92 &  4.7 \\
8 & 58918 & 58872.5 & 58964.4 & 14 & 61.909033102(3)   & $-1.99765(1)$  & 1.4(1)  &   22(2) & 138.4 &  6.9 \\
\hline
\multicolumn{11}{l}{$^a$\,$\nuddot$ is fixed at $10^{-20}\mbox{ Hz s$^{-2}$}$
(see text). $^b$No fit performed (see text).}
\end{tabular}
\end{table*}

The timing solutions determined above provide an initial estimate of
glitch sizes.
Each glitch is measured more precisely by fitting a timing model
(see, e.g., \citealt{edwardsetal06}) that
includes glitch parameters $\Delta\phi$, $\Delta\nus$, $\Delta\nudot$,
and $\Delta\nuddot$ to the set of TOAs that immediately precedes the
glitch and the set of TOAs that immediately follows the glitch
(e.g., TOAs from segments 0 and 1 are fit to obtain parameters of
glitch 1);
like glitches detected using \RXTE\ \citep{antonopoulouetal18},
an exponential recovery-term is not needed to describe our measured
glitches (see also Section~\ref{sec:properties}).
However, for glitch 1, $\Delta\nudot$ is not varied during the fitting
but is calculated by comparing $\nudot$ from segments 0 and 1 at the
glitch epoch; its uncertainty is the uncertainty of $\nudot$ in segment 1.
Because $\nudot$ cannot be varied for segment~1, the uncertainty of
$\Delta\nudot$ for glitch~2 is set to the uncertainty of $\nudot$
for segment~1.
For glitches 1 and 5, $\Delta\nuddot$ is set to zero during the fitting
because $\nuddot$ is not determined in segments 1 and 5 (see above);
for the same reason, $\Delta\nuddot$ for glitch 2 (and 6) is simply
the change from the fixed $\nuddot=10^{-20}\mbox{ Hz s$^{-2}$}$ in
segment 1 (and 5) to the measured $\nuddot$ in segment 2 (and 6).
Glitch epochs are set at the centre of the interval between the
last TOA before the glitch and the first TOA after the glitch, and the
uncertainty is set at half this interval.
The results of these timing model fits are given in Table~\ref{tab:glitch}.

\begin{table*}
\centering
\caption{\psr\ glitch parameters.
Number in parentheses is 1$\sigma$ uncertainty in last digit.}
\label{tab:glitch}
\begin{tabular}{clcccc}
\hline
Glitch & Glitch epoch & $\Delta\phi$ & $\Delta\nus$ & $\Delta\nudot$ & $\Delta\nuddot$ \\
 & (MJD) & (cycle) & ($\mu$Hz)$\quad$ & ($10^{-13}$ Hz s$^{-1}$) & ($10^{-20}$ Hz s$^{-2}$) \\
\hline
1 & 58083(25) & $-0.016(8)$ & 16.132(2) & $-1.5(3)$  & --- \\
2 & 58152(11) & $ 0.47(1)$  & 36.035(6) & $-1.6(3)$  & $-0.44(1)^a$ \\
3 & 58363(14) & $ 0.17(5)$  &  7.83(5)  & $-2.3(4)$  & $ 5(1)$ \\
4 & 58424(2)  & $-0.35(23)$ & 25.3(3)   & $-2(1)$    & $-5(2)$ \\
5 & 58566(5)  & $-0.32(2)$  &  9.21(2)  & $-0.89(5)$ & --- \\
6 & 58637(8)  & $ 0.03(2)$  & 26.99(1)  & $-0.86(4)$ & $-0.12(3)^a$ \\
7 & 58807(3)  & $ 0.31(2)$  &  7.57(3)  & $-2.2(3)$  & $ 5(1)$ \\
8 & 58868(5)  & $ 0.06(6)$  & 24.04(8)  & $-2.4(5)$  & $-4(1)$ \\
\hline
\multicolumn{6}{l}{$^a$In the preceding segment, $\nuddot$ is fixed at
$10^{-20}\mbox{ Hz s$^{-2}$}$ due to a low number of TOAs (see} \\
\multicolumn{6}{l}{Table~\ref{tab:data}).  $\Delta\nuddot$ is the
difference between this fixed value and $\nuddot$ in the next segment.}
\end{tabular}
\end{table*}

\begin{figure}
\begin{center}
\includegraphics[width=\columnwidth]{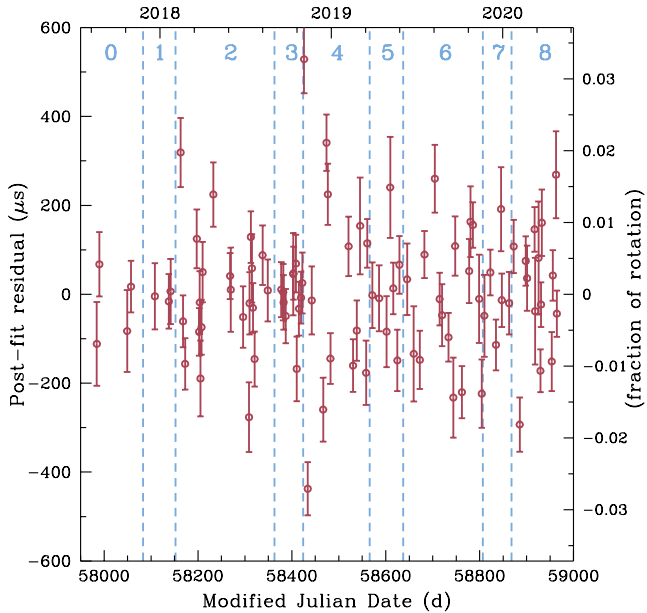}
\caption{
Timing residual for each merged observation after accounting for
the best-fit timing model, with left axis in $\mu$s and right axis in
fraction of rotation cycle.
Errors are 1$\sigma$ uncertainty.
Segments are labeled by numbers and separated by the occurrence
of a glitch, each of which is denoted by a vertical dashed line.
}
\label{fig:trms}
\end{center}
\end{figure}

The glitch parameters determined above are used together to create
a single timing solution for all measured TOAs.
The reference time is set at the centre of segment~0, and the reference
$\nus$, $\nudot$ and $\nuddot$ are set at corresponding values for segment~0.
The resulting best-fit timing model has a RMS fit residual of
$141\mbox{ $\mu$s}$, i.e., $<0.9$~percent of the rotation cycle of \psr.
The glitch parameters obtained by this subsequent fit agree within errors
to those measured by individual fits above (see Table~\ref{tab:glitch}).
Figure~\ref{fig:trms} shows the residual of each TOA after removing
the final best-fit timing model.

\subsection{Long and short-term braking indices} \label{sec:braking}

The spin-down behavior of a pulsar can be described by the braking
index, which is defined as $n\equiv\nu\nuddot/\nudot^2$ and follows
from characterizing the spin-down rate as a power law $\nudot\propto-\nu^n$.
In the case of \psr, its spin-down rate is measured to be
increasing ($\nudot$ becoming more negative) over time,
with \citet{antonopoulouetal18} finding
$\nuddot=(-7.7\pm0.3)\times10^{-22}\mbox{ Hz s$^{-2}$}$ and $n=-1.22\pm0.04$
and \citet{ferdmanetal18} finding
$\nuddot=(-8.2\pm0.3)\times10^{-22}\mbox{ Hz s$^{-2}$}$ and $n=-1.28\pm0.04$.

We perform a simple linear fit of $\nudot$ over time, which yields
$n=-1.32\pm0.02$ for only \RXTE\ data
(taken from Table~1 of \citealt{antonopoulouetal18})
and $n=-1.4\pm0.2$ for only \NICER\ data (see Table~\ref{tab:data});
1$\sigma$ errors are obtained by increasing all individual $\nudot$ errors
by a single factor such that the fit produces a $\chi^2/\mbox{dof}=1$.
A fit of all $\nudot$ values gives
$\nuddot=(-8.00\pm0.08)\times10^{-22}\mbox{ Hz s$^{-2}$}$, which is shown
in Figure~\ref{fig:tf1}, and long-term braking index $n=-1.25\pm0.01$.

\begin{figure}
\begin{center}
\includegraphics[width=\columnwidth]{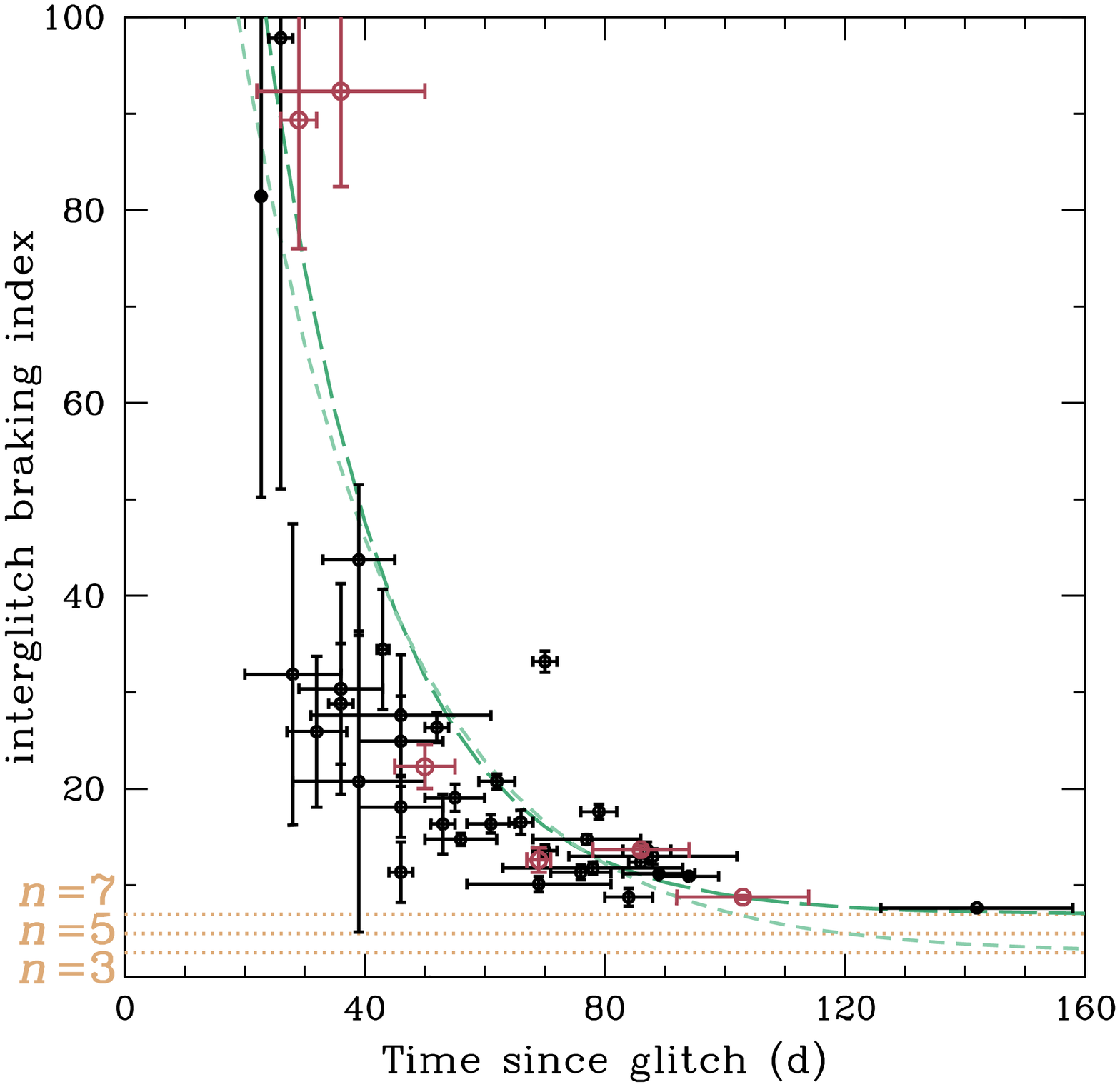}
\caption{
Braking index $\nig$ calculated from spin parameters of each segment
between glitches as a function of time since last glitch.
Large and small circles denote \NICER\ and \RXTE\ values, respectively,
with the latter taken from Tables~1 and 2 of \citet{antonopoulouetal18}.
Errors in $\nig$ are 1$\sigma$ uncertainty.
Horizontal dotted lines indicate braking index $n=3,5,7$, which are
expected for pulsar spin-down due to electromagnetic dipole radiation,
gravitational-wave emitting ellipticity, and
gravitational-wave emitting r-mode oscillation, respectively
(see Section~\ref{sec:gw}).
Short and long-dashed lines indicate exponential decay to $n=3$
with best-fit timescale of 26~d and to $n=7$ with best-fit timescale
of 20~d, respectively.
}
\label{fig:tnig}
\end{center}
\end{figure}

The long-term spin-down behavior of \psr\ is in sharp contrast
to the behavior over short interglitch intervals.
Between individual glitches, the interglitch braking index $\nig$
is generally non-negative and much greater than the canonical value of 3
for spin-down by electromagnetic dipole radiation (see Table~\ref{tab:data}
and \citealt{antonopoulouetal18}; see also \citealt{middleditchetal06}).
Furthermore, \citet{anderssonetal18} show that, when measuring braking
indices as a function of time since the preceding glitch, $\nig$
tends to decrease toward an asymptotic value.
Figure~\ref{fig:tnig} shows \NICER\ values alongside \RXTE\ values.
It is likely that large braking indices determined at short times
after a glitch reflect the impact of the glitch on the spin-down
behavior of the pulsar.
This is borne out in modeling of post-glitch relaxation, with
an exponential timescale of 17--34~d and an asymptotic $\nig\approx 7$
\citep{anderssonetal18,antonopoulouetal18,ferdmanetal18}.
Observationally, a more reliable measure of spin-down behavior with
less glitch contamination can be obtained by determining braking
indices at long times after a large glitch.
While our current \NICER\ dataset contributes only seven new interglitch
braking indices, we see that \NICER\ values of $\nig$ follow the
same trend as those from \RXTE\ and are highlighted by interglitch
segment 2 which follows $103\pm11\mbox{ d}$ after the second largest
glitch observed in \psr\ (glitch 2; see Table~\ref{tab:glitch}).
We perform a simple fit with the form $\nig=\nig^\infty+n_0e^{-t/\tig}$,
where $n_0$ and $\tig$ are fit parameters and $\nig^\infty$ is the asymptotic
value of the braking index which we take to be either 3, 5, or 7;
fits yield decay timescales of 19--44~d, with a longer timescale
for a lower braking index (see Figure~\ref{fig:tnig}).
While assuming an asymptotic $\nig^\infty=7$ leads to a better fit
than assuming $\nig^\infty=3$ or 5, the frequent occurrence of
glitches may be preventing measurements of $\nig<7$.
We discuss implications of braking indices of 5 and 7 in Section~\ref{sec:gw}.

\section{Glitches} \label{sec:glitch0}

\subsection{Glitch properties} \label{sec:properties}

The detection of these 8 glitches, with parameters given in
Table~\ref{tab:glitch}, in 2.7~yr of observation yields a glitch
rate of 3.0~yr$^{-1}$. Thus \psr\ continues to glitch at the same rate
as during the \RXTE\ era, when either 42 glitches \citep{ferdmanetal18}
or 45 glitches \citep{antonopoulouetal18} are found in 12.95~yr of
\RXTE\ data at a rate of either 3.24~yr$^{-1}$ or 3.47~yr$^{-1}$,
respectively.

\begin{figure}
\begin{center}
\includegraphics[width=\columnwidth]{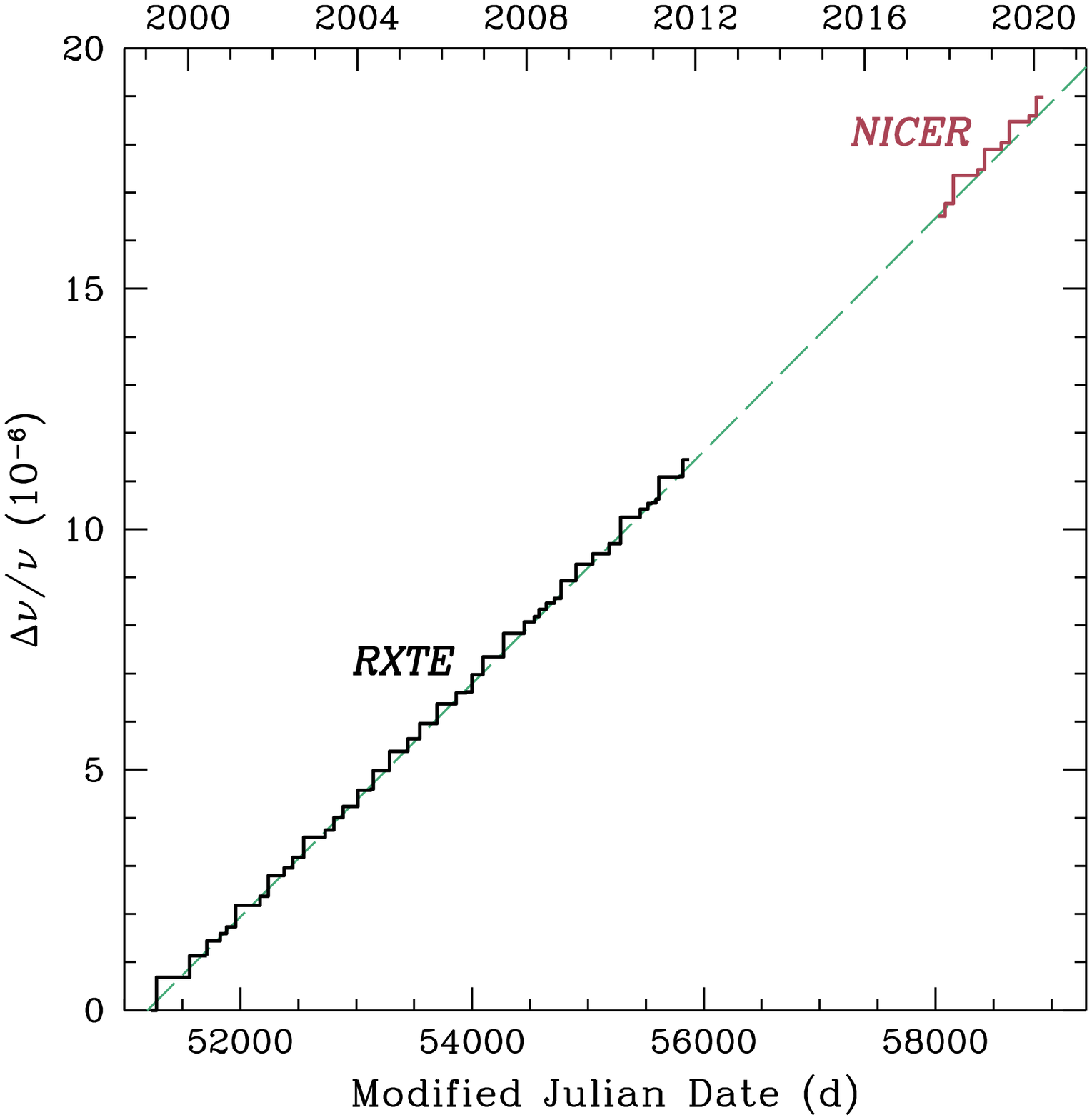}
\caption{
Fractional glitch magnitude $\Delta\nus/\nus$ shown as a 
cumulative sum over each previous glitch.
\RXTE\ values are from Table~2 of \citet{antonopoulouetal18},
Dashed line indicates a line with a slope of
$8.84\times 10^{-7}\mbox{ yr$^{-1}$}$, which is the glitch activity
$\Ag\equiv\sum_i(\Delta\nus/\nus)_i/\tobs$ from \RXTE\ data, where
$\tobs$ is time over which the pulsar is monitored;
note that $\Ag=8.88\times 10^{-7}\mbox{ yr$^{-1}$}$ for the combination
of \RXTE\ and \NICER\ glitches.
\NICER\ values are offset by $\Delta\nus/\nus=16.5\times10^{-6}$,
i.e., value of dashed line at the epoch of \NICER\ segment 0 at MJD 58020.
}
\label{fig:df0cum}
\end{center}
\end{figure}

Glitching activity of a pulsar can also be characterized by the parameter
$\Ag\equiv\sum_i(\Delta\nus/\nus)_i/\tobs$, where the summation is over
each glitch $i$ and $\tobs$ is time over which the pulsar is
monitored \citep{mckennalyne90}.
For glitches detected using \RXTE\ and $\tobs=12.95\mbox{ yr}$,
the results of \citet{antonopoulouetal18} yield
$\sum_i\Delta\nus_i=(709.5\pm0.8)\mbox{ $\mu$Hz}$ and
$\Ag=(8.84\pm0.01)\times 10^{-7}\mbox{ yr$^{-1}$}$,
while the results of \citet{ferdmanetal18} yield
$\sum_i\Delta\nus_i=(707.4\pm0.8)\mbox{ $\mu$Hz}$ and
$\Ag=(8.81\pm0.01)\times 10^{-7}\mbox{ yr$^{-1}$}$.
For glitches detected using \NICER\ and $\tobs=2.7\mbox{ yr}$,
values from Table~\ref{tab:glitch} give
$\sum_i\Delta\nus_i=(153.1\pm0.3)\mbox{ $\mu$Hz}$ and
$\Ag=(9.22\pm0.02)\times 10^{-7}\mbox{ yr$^{-1}$}$.
Figure~\ref{fig:df0cum} plots the cumulative fractional glitch
magnitude $\Delta\nus/\nus$ over the \RXTE\ and \NICER\ eras.
Combining \RXTE\ and \NICER\ glitches produces an activity parameter
$\Ag=(8.88\pm0.01)\times 10^{-7}\mbox{ yr$^{-1}$}$.
We note that \citet{ferdmanetal18} find an indication that $\Ag$
decreases during the time \RXTE\ is monitoring \psr.
However, we find that $\Ag$ during the \NICER\ era is similar to or
even greater than $\Ag$ measured during the \RXTE\ era.

\begin{figure}
\begin{center}
\includegraphics[width=\columnwidth]{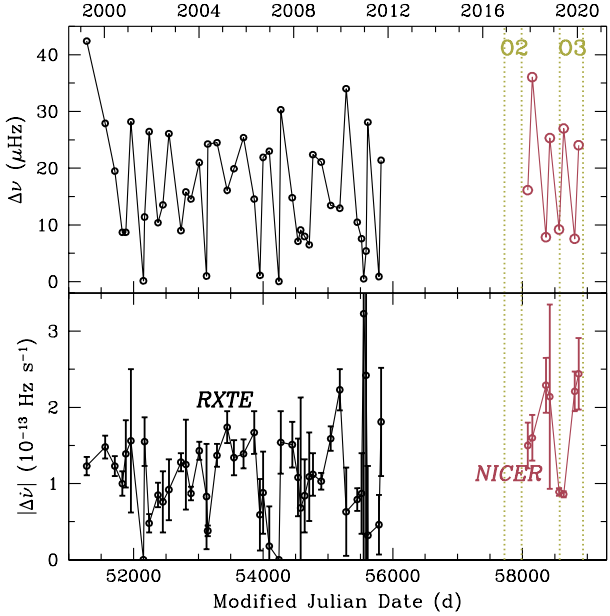}
\caption{
Glitch $\Delta\nus$ (top) and $\Delta\nudot$ (bottom) as functions of time.
Large and small circles denote \NICER\ and \RXTE\ values, respectively,
with the latter taken from Table~2 of \citet{antonopoulouetal18}.
Errors in bottom panel are 1$\sigma$ uncertainty, while errors are
not shown in top panel because they are smaller than the circles.
Vertical dotted lines denote the
start (MJD 57722, 2016 November 30) and
end (MJD 57990, 2017 August 25) of the second observing run (O2) and
start (MJD 58574, 2019 April 1) and
end (MJD 58935, 2020 March 27) of the third observing run (O3) of LIGO/Virgo.
}
\label{fig:glitch}
\end{center}
\end{figure}

Figure~\ref{fig:glitch} shows measured $\Delta\nus$ and $|\Delta\nudot|$.
\citet{middleditchetal06} argue that there exists a hard upper
limit of $|\Delta\nudot|=1.5\times 10^{-13}\mbox{ Hz s$^{-1}$}$
based on the first 23 glitches detected using \RXTE.
However, analysis of the next 22 glitches reveal several
exceeding this limit, with three even having
$|\Delta\nudot|>2\times 10^{-13}\mbox{ Hz s$^{-1}$}$ \citep{antonopoulouetal18};
analysis by \citet{ferdmanetal18} find somewhat lower $|\Delta\nudot|$
but still several exceeding the proposed limit.
While \citet{antonopoulouetal18} indicate such large $|\Delta\nudot|$
should be rare (based on 3 out of 45), \NICER\ data reveals 4 out of 8
glitches with $|\Delta\nudot|>2\times 10^{-13}\mbox{ Hz s$^{-1}$}$,
although their uncertainties are large enough to reach below this limit.

The 8 glitches detected using \NICER\ show potentially correlated
properties, in addition to the same glitch size--time to next glitch
correlation observed in \RXTE\ glitches (see Section~\ref{sec:predict}).
\NICER\ glitches appear in pairs, with each pair consisting first of
a smaller glitch ($\Delta\nus\approx8-16\mbox{ $\mu$Hz}$) and then a
larger glitch ($\Delta\nus\approx24-36\mbox{ $\mu$Hz}$) but each glitch
in a pair having similar $\Delta\nudot$ and $|\Delta\nuddot|$
(and $\Delta\nuddot>0$ for the first glitch in the pair,
then $\Delta\nuddot<0$ for the second glitch in the pair),
albeit with large uncertainties
(see Table~\ref{tab:glitch} and Figure~\ref{fig:glitch}).
Each pair occurs quasi-periodically
(as is evident from, e.g., Figure~\ref{fig:f0resid}),
with a period of $\sim 210-270\mbox{ d}$ (in $\Delta\nu$) that simply
reflects the glitch size--time to next glitch correlation and
$\sim 440-480\mbox{ d}$ (in $\Delta\nudot$).
These behaviors do not seem to occur in the \RXTE\ era, and
it will be interesting to see whether this continues with future
monitoring and detection of glitches using \NICER.

For the three largest glitches (glitches 2, 4, and 6), we add an
exponential term in the timing model to look for a potential
exponential recovery associated with each glitch.
No evidence of exponential recovery for glitch 6 is found,
but there is marginal evidence for glitches 2 and 4
(with $\chi^2/\mbox{dof}$ decreasing from 4.1 to 2.6 and from 12 to 6.9,
respectively), each with a timescale of $\sim$5~d.
This is in contrast to the 20~d timescale of the largest \RXTE\ glitch
\citep{antonopoulouetal18}.

\subsection{Glitch predictability} \label{sec:predict}

\begin{figure}
\begin{center}
\includegraphics[width=\columnwidth]{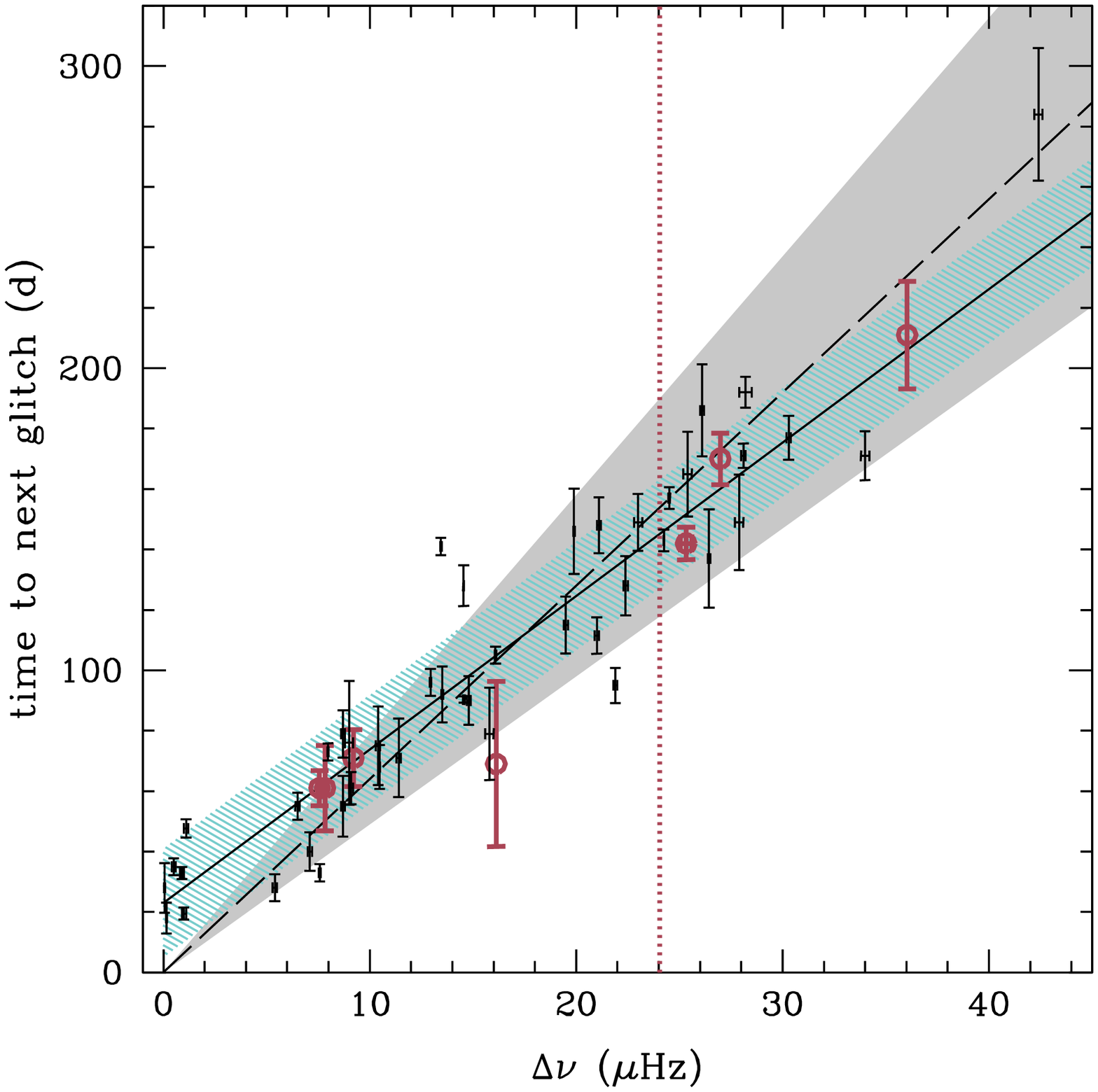}
\caption{
Time to next glitch as a function of glitch size $\Delta\nus$.
Circles with errors are \NICER\ values, and points with errors are
\RXTE\ values, which are from Table~2 of \citet{antonopoulouetal18}.
Vertical dotted line indicates the size of glitch 8, which is the
most recent \NICER-detected glitch and for which the time to next
glitch is not known yet.
Errors in $\Delta\nus$ are 1$\sigma$ uncertainty.
Solid and dashed lines show a linear fit of \NICER\ and \RXTE\ data with
time to next glitch $=6.4\mbox{ d }(\Delta\nus/\mbox{$\mu$Hz})$
and
time to next glitch $=23\mbox{ d}+5.08\mbox{ d }(\Delta\nus/\mbox{$\mu$Hz})$,
respectively,
while shading shows correlation regions which fit 68~percent of glitches
(see Section~\ref{sec:predict}).
}
\label{fig:tdf0}
\end{center}
\end{figure}

\psr\ is unique in how well its glitch sizes $\Delta\nus$ correlate
with time to the next glitch
\citep{middleditchetal06,antonopoulouetal18,ferdmanetal18}.
Figure~\ref{fig:tdf0} shows this correlation, with \NICER\ glitches
from Table~\ref{tab:glitch} and \RXTE\ glitches from
\citet{antonopoulouetal18}.
We perform two fits to model the correlation: one linear model
is constrained such that the time to next glitch is zero when
$\Delta\nus=0$ and a second linear model that allows for
non-zero time to next glitch when $\Delta\nus=0$.
One might argue that the former is more physically-motivated.
However, Figure~\ref{fig:tdf0} clearly shows that there is a delay
even for small glitches, which could suggest that there is a minimum
time to next glitch or that the correlation has a different slope or
becomes non-linear at small glitch sizes.
Results of the fits to all \NICER\ and \RXTE\ glitches are
time to next glitch $=6.4\mbox{ d }(\Delta\nus/\mbox{$\mu$Hz})$ and
time to next glitch
$=(23\pm1)\mbox{ d }+(5.08\pm0.07)\mbox{ d }(\Delta\nus/\mbox{$\mu$Hz})$,
and both fit results are plotted in Figure~\ref{fig:tdf0}.
Fits to only \NICER\ or only \RXTE\ glitches yield very similar
results, and the above results also closely match those obtained
by \citet{antonopoulouetal18,ferdmanetal18} using only \RXTE\ glitches.

While these best-fit results indicate a formal uncertainty in time to
next glitch of only a few days, comparison of the two fit correlations
can produce a difference of $\sim 10\mbox{ d}$ in predicted time, and
observed values show significant scatter around the best-fit correlations.
To obtain a more reliable estimate of the uncertainty in predicted
time to next glitch, we expand the parameter space covered by the
best-fit correlation until 68~percent of measured glitches (with a
follow-up glitch) fall within the expanded region.
In particular, for the fit that is constrained at $\Delta\nus=0$,
we increment symmetrically the slope around the best fit value until
35 of 51 glitches lie within the two lines defined, i.e.,
time to next glitch $=(6.4\pm1.5)\mbox{ d }(\Delta\nus/\mbox{$\mu$Hz})$.
For the fit that is unconstrained at $\Delta\nus=0$, we increment the
normalization rather than the slope and find the two lines defined by
time to next glitch
$=(23\pm18)\mbox{ d }+5.08\mbox{ d }(\Delta\nus/\mbox{$\mu$Hz})$
encompass 68~percent of glitches.
Each of these are illustrated in Figure~\ref{fig:tdf0}.
From these results, the actual uncertainty in time to next glitch
is $\sim \pm 20\mbox{ d}$.

\subsection{Glitch-induced emission variability} \label{sec:variability}

We present a limited search for potential spectral and pulse
profile changes associated with a glitch
because the large non-imaging field-of-view of \NICER\ is not ideal
for studying emission from a weak point source embedded within
bright diffuse emission, such as the case of \psr.
Note that \citet{ferdmanetal18} look for but do not find
glitch-associated flux and pulse profile changes using the
\RXTE\ dataset
(see also, e.g., \citealt{palfreymanetal18} for a radio search during
a Vela pulsar glitch,
\citealt{rayetal19} for a X-ray search during three spin-down glitches
of NGC~300 ULX-1,
and \citealt{fengetal20} for a X-ray polarization search during a
Crab pulsar glitch).
Here, we identify \NICER\ observations that are nearest in time
before and after glitch 4 (see Table~\ref{tab:glitch}) on MJD 58424
(2018 November 2) and
compare spectra and pulse profiles to see if there are any clear
signs of variability that could be associated with the glitch.
Note that glitch 7 has the next nearest in time observations,
about three days on either side of the glitch.

For spectral analysis, ObsID 1020100312 (MJD 58421; October 30) and
1020100313 (MJD 58422; October 31) are taken about two to three days
before the glitch and ObsID 1020100314 (MJD 58425; November 3) and
1020100315 (MJD 58426; November 4) are taken about one to two days
after the glitch.  Data are processed using \texttt{nicerl2}
in NICERDAS with standard filtering options.
We apply a barycentric correction, extract spectra from clean event data,
subtract a background generated using the \texttt{nibackgen3C50} tool,
and add and bin spectra using \texttt{mathpha} and \texttt{grppha},
respectively.
We obtain net exposures of 6.8~ks and 5.3~ks from the above specified
pre and post-glitch observations, respectively.
A comparison of the spectra results in $\chi^2/\mbox{dof}=552/314$
and no distinctive features or clearly visible differences.
We also fit each spectrum at 1--5~keV with an absorbed power law model
and obtain model parameters before and after the glitch that are
consistent within uncertainties,
i.e., absorption $N_{\rm H}=(2.7\pm0.6)\times 10^{21}\mbox{ cm$^{-2}$}$
versus $(2.2\pm0.6)\times 10^{21}\mbox{ cm$^{-2}$}$,
power law index $\Gamma=2.7\pm0.1$ versus $2.5\pm0.1$,
and unabsorbed flux $=(6.0\pm0.2)\times 10^{-12}\mbox{ erg s$^{-1}$ cm$^{-2}$}$
versus $(6.1\pm0.2)\times 10^{-12}\mbox{ erg s$^{-1}$ cm$^{-2}$}$;
errors are at 90~percent confidence.
Imaging spectra of the supernova remnant using \Chandra\ and \XMM\
\citep{chenetal06,kuiperhermsen15}
indicate higher $N_{\rm H}$ ($=6\times 10^{21}\mbox{ cm$^{-2}$}$)
but comparable power law index ($\Gamma\sim 2.4$)
and flux ($\approx1\times 10^{-11}\mbox{ erg s$^{-1}$ cm$^{-2}$}$)
to our spectral results.
Therefore we conclude that our spectra are dominated by remnant
emission, and pulsed emission from \psr\ would be difficult to extract,
which limits our ability to measure glitch-induced spectral variability.

\begin{figure}
\begin{center}
\includegraphics[width=\columnwidth]{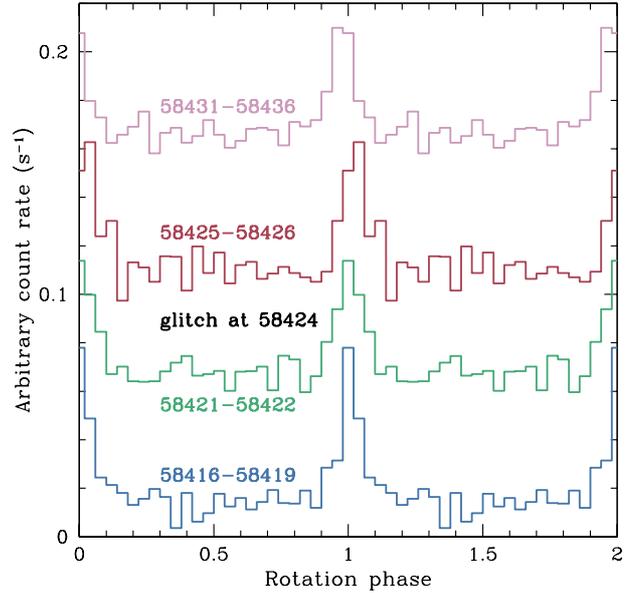}
\caption{
Pulse profiles of \psr\ in the 1--7~keV band, folded at the ephemeris
from Tables~\ref{tab:data} and \ref{tab:glitch}, from earliest date at
the bottom to latest date at the top.
Two rotation cycles are shown, with 25 bins per cycle, and
each profile is arbitrarily shifted in count rate for clarity.
Labels indicate MJD of data used to construct profile and epoch
of glitch 4 (at MJD 58424).
}
\label{fig:profile}
\end{center}
\end{figure}

For pulse profile analysis, our data reduction procedure is as described
in Section~\ref{sec:data} and yields exposures of 7.2~ks and 4.7~ks for
the pre and post-glitch observations described above, respectively.
We also consider observations from a somewhat larger range of dates
around the glitch epoch.  In particular, ObsID 1020100309--1020100311
(MJD 58416--58419; October 25--28) are from before the glitch and
ObsID 1020100317--1020100321 (MJD 58431--58436; November 9--14) are
from after the glitch, and merged observations from these ObsIDs
have exposures of 7.5~ks and 8.4~ks, respectively.
Figure~\ref{fig:profile} shows a comparison of the pulse profiles from
before and after the glitch.
Like the spectra, there are no clear significant differences.
Subtracting from each pulse profile an average pulse profile,
constructed from the above observations, we find residuals that are
constant within uncertainties and a shift of one phase bin.

\section{Discussions} \label{sec:discuss}

In this work, we present results of our ongoing campaign of monitoring
and timing the pulsar \psr\ using \NICER.
We are able to obtain a rotation phase-connected timing model for
the evolution of the pulsar spin frequency $\nus$ during the first
2.7~yr of \NICER\ observations.
In agreement with timing results using \RXTE\ from 1999--2011,
we measure a long-term braking index $n=-1.25\pm0.01$ and a
short-term braking index between glitches that seems to relax
over time towards a value of $\nig=7$ or possibly lower.
We measure 8 glitches, with similar properties as \RXTE\ glitches, and
the rate is in line with the 42 or 45 glitches measured in 13 years
of \RXTE\ data.
The 8 glitches also show a pairing/periodicity whose significance
will be tested by future observations.
We obtain a more reliable estimate of the uncertainty in the observed
correlation between glitch size and time to next glitch.
Finally, for the glitch that was most closely observed in time,
i.e., within $\pm 2\mbox{ d}$, we search for but do not see clear
evidence of spectral or pulse profile changes that could be associated
with the glitch.

We provide here limited discussions of implications of our monitoring
results and newly detected glitch activity.
This is because much of the timing and glitch behaviors seen
using the current \NICER\ dataset is in good agreement with the
behaviors seen using the \RXTE\ dataset, and implications of the
latter dataset are described extensively in many previous works
(see below).

\subsection{Superfluid and crust moment of inertia and pulsar mass}
\label{sec:sf}

The mechanism that produces large spin-up glitches like those seen
in the Vela pulsar \citep{dodsonetal07,palfreymanetal18} and \psr\ is
thought to be a sudden transfer of angular momentum from a rapidly rotating
superfluid that permeates the neutron star inner crust to the rest
of the star, which is slowing down from electromagnetic radiation braking
\citep{andersonitoh75,haskellmelatos15,graberetal17}.
\citet{linketal99} provide strong support for this theory by first
showing that the fractional moment of inertia ($\Isf/I$) of the
superfluid angular momentum reservoir is related to a pulsar's glitch
activity via $\Isf/I\ge2\tauc\Ag$, where $I$ is total moment
of inertia and $\tauc=-\nu/2\nudot$ is pulsar characteristic age.
They then show that the glitch activity of pulsars like Vela gives
$\Isf/I\gtrsim 0.01$, and this approximately matches the theoretical
fractional moment of inertia of a neutron star's crust
$I_{\rm crust}/I$.  For \psr, we find $2\tauc\Ag=0.00874$,
which agrees with that found in previous works
\citep{middleditchetal06,antonopoulouetal18,ferdmanetal18}.
More recently, calculations by \citet{anderssonetal12,chamel13}
indicate that the above relation is underestimated after accounting
for superfluid entrainment and should instead be
$\Isf/I\ge2\tauc\Ag\mneff/\mn$, where $\mneff$ and $\mn$ are
averaged effective neutron mass and neutron mass, respectively
(and $\mneff/\mn\approx 4.2$; \citealt{chamel12}), and as a result,
the angular momentum required by a glitch is more than the crust
can provide.
One solution proposed by \citet{anderssonetal12} is that the
superfluid component in the crust extends into the core.
Subsequently, \citet{hoetal15,hoetal17} show that, by combining a
pulsar's glitch activity with a measurement of true age or surface
temperature, one can obtain valuable constraints on properties of the
superfluid and even measure the mass of the pulsar;
for the $\Ag$ and estimated age of \psr, the pulsar's mass turns
out to be much higher than the canonical $1.4 M_\odot$.
Very recently, \citet{saulsetal20} suggest superfluid entrainment
may not be as strong as found by \citet{chamel12,chamel17}.
More work is needed to resolve the issue.

\subsection{Braking index and GW implications} \label{sec:gw}

As described in Section~\ref{sec:braking}, we determine a long-term
spin-rate change $\nuddot=(-8.00\pm0.08)\times10^{-22}\mbox{ Hz s$^{-2}$}$
from spin-rate $\nudot$ measurements of \psr\ over the past 21 years
(Figure~\ref{fig:tf1}),
which indicates an increasing $|\nudot|$ over time, unlike for most pulsars.
The corresponding braking index $n\equiv\nu\nuddot/\nudot^2=-1.25\pm0.01$
is significantly lower than the canonical value of 3, which is implied
for a pulsar whose rotational energy loss rate is purely due to
electromagnetic dipole radiation \citep{shapiroteukolsky83}.
The braking index of \psr\ is the lowest precisely measured value
(cf. J1738$-$2955 has $n=-70\pm40$ and J1833$-$0827 has $n=-15\pm2$;
\citealt{parthasarathyetal20}).
Glitches could contribute to lowering the braking index, as possibly
indicated by the prevalence of low $n$ for glitching pulsars
(\citealt{espinozaetal17}; see also \citealt{ho15}).
Various other ideas and models that can produce a braking index below 3
have been proposed, e.g.,
a changing moment of inertia due to superfluidity \citep{hoandersson12},
evolution of magnetic field orientation
\citep{middleditchetal06,lyneetal15,johnstonkarastergiou17}
or magnetic field strength \citep{romani90},
as demonstrated for the case of \psr\ \citep{gourgouliatoscumming15,ho15},
and particle winds
\citep{michel69,micheltucker69,tongetal13}.

Meanwhile, measurements of the braking index of \psr\ over shorter
timescales of tens of days between glitches yields much larger values
(Table~\ref{tab:data}).
While the long-term spin evolution is clearly affected by glitches,
the strength of the effect is unclear.
Regardless, one can see that the post-glitch spin behavior appears
to relax at long times after a glitch to a braking index value near 7
or even lower (Figure~\ref{fig:tnig}; see also \citealt{anderssonetal18}),
which is supported by a fit of post-glitch spin-rate with a model that
assumes changes on an exponential timescale (see Section~\ref{sec:braking}).

In addition to mechanisms that can lead to a braking index less than 3,
such as those described above, it is well-known that neutron stars
emitting GWs can have a braking index greater than 3.
For example, a neutron star with a quadrupolar mass deformation,
characterized by an ellipticity $\varepsilon\equiv|I_{xx}-I_{yy}|/I_{zz}$,
where $I_{xx}$, $I_{yy}$, and $I_{zz}$ are triaxial components
of the stellar moment of inertia, can emit GWs at a frequency
$\nugw$ ($=2\nus$) with a strain amplitude
\begin{eqnarray}
h_0 &=& \frac{16\pi^2G}{c^4}\frac{\varepsilon I_{zz}}{d}\nus^2 \nonumber\\
&=& 4.23\times 10^{-26}\left(\frac{\varepsilon}{10^{-5}}\right)
 \left(\frac{10 \mbox{ kpc}}{d}\right)\left(\frac{\nus}{100\mbox{ Hz}}\right)^2
\label{eq:h0}
\end{eqnarray}
(see, e.g., \citealt{abbottetal19c}).
The GW strain amplitude $h_0$, which is either measured or has an upper
bound in the case of non-detection, can be compared to the ``spin-down limit''
strain amplitude $\hsd$, such that a constraint on a pulsar's
ellipticity is obtained when $h_0/\hsd<1$.  The spin-down limit
\begin{eqnarray}
\hsd &=& \left(-\frac{5G}{2c^3}\frac{I_{zz}}{d^2}\frac{\nudot}{\nu}\right)^{1/2}
 \nonumber\\
&=& 8.06\times 10^{-26}\left(\frac{10 \mbox{ kpc}}{d}\right)
 \left(\frac{100\mbox{ Hz}}{\nus}\right)^{1/2}
 \left(\frac{-\nudot}{10^{-10}\mbox{ Hz s$^{-1}$}}\right)^{1/2}, \label{eq:hsd}
\end{eqnarray}
is determined by assuming that a neutron star's rotational energy loss is
due entirely to GW quadrupolar emission and implies a braking index $n=5$.
As GW searches become more sensitive, such that the measured $h_0$
decreases, an improving constraint on ellipticity is
\begin{equation}
\varepsilon = 1.91\times 10^{-5}\left(\frac{100\mbox{ Hz}}{\nus}\right)^{5/2}
 \left(\frac{-\nudot}{10^{-10}\mbox{ Hz s$^{-1}$}}\right)^{1/2}
 \left(\frac{h_0}{\hsd}\right).
\end{equation}

The latest LIGO/Virgo searches of \psr\ find an upper limit of
$h_0/\hsd=1.92$ using
GW data from the second observing run (O2), which collected data
from 2016 November 30 to 2017 August 25 \citep{abbottetal19a}.
Because the only timing model of \psr\ available at the time of the
GW search is that derived from \RXTE\ observations from 1999--2011,
the search by \citet{abbottetal19a} could only be conducted in a
frequency band around the frequency expected for \psr\ and could not
account for glitches \citep{ashtonetal17,keiteletal19}.
A contemporaneous timing model, such as that determined here using \NICER,
enables the most sensitive targeted searches.
Estimating the $h_0$ sensitivity upper limit of O2 data at
$\nugw=124\mbox{ Hz}$ from Figure~2 of \citet{abbottetal19c},
such a targeted search could yield $h_0/\hsd<0.4$,
which would set the limit $\varepsilon<4\times10^{-5}$
and limit the percentage of rotational energy loss that is due to
GW emission to $<20$~percent.
We also note that, while \psr\ is much more distant than pulsars in
the Milky Way, the decrease in signal strength
[e.g., equation~(\ref{eq:h0})] is partly compensated by its higher
GW frequency, which is in the most sensitive range of ground-based
GW detectors.

An alternative GW emission mechanism is that due to fluid motions in
a neutron star.  For example, an active r-mode oscillation can produce
GWs \citep{andersson98,friedmanmorsink98,anderssonkokkotas01},
and GW searches for r-modes in pulsars are ongoing
(see, e.g., \citealt{abbottetal19b}).
The GW frequency in the case of r-modes is $\nugw\sim(4/3)\nus$
[or $\nugw\sim(1.4-1.6)\nus$ after accounting for relativistic corrections;
\citealt{anderssonetal14,idrisyetal15,jasiulekchirenti17,carideetal19}],
and the theoretical strain amplitude equivalent to equation~(\ref{eq:h0}) is
\begin{equation}
h_0 = 5.36\times 10^{-26}\left(\frac{\alpha}{10^{-2}}\right)
 \left(\frac{10 \mbox{ kpc}}{d}\right)\left(\frac{\nus}{100\mbox{ Hz}}\right)^3,
\end{equation}
while the spin-down limit $\hsd$ is 3/2 that given by equation~(\ref{eq:hsd})
\citep{owen10} and leads to a limit on the r-mode amplitude
\begin{equation}
\alpha = 0.02\left(\frac{100\mbox{ Hz}}{\nus}\right)^{7/2}
 \left(\frac{-\nudot}{10^{-10}\mbox{ Hz s$^{-1}$}}\right)^{1/2}
 \left(\frac{h_0}{\hsd}\right).
\end{equation}

What makes r-modes of special relevance to \psr\ is that, if the spin
evolution of a neutron star is dominated by GW emission from an active
r-mode, then the resulting braking index is 7 \citep{owenetal98},
although it can also be less than 7 \citep{alfordschwenzer14}.
As we show in Section~\ref{sec:braking}
(see also \citealt{anderssonetal18,ferdmanetal18}),
the braking index between glitches of \psr\ seems to relax over
time towards a value of 7, as effects of the glitch on long-term
spin evolution decrease.
If correct, this suggests an active r-mode and the spin evolution
of \psr\ is determined by GW emission.
Such a situation is examined in \citet{anderssonetal18},
who conclude that an active r-mode is marginally possible,
especially with uncertainties and discrepancies in our
understanding of r-modes and their impact on observations
(see, e.g., \citealt{hoetal19}, and references therein).

Narrow-band searches for GWs generated by r-modes in \psr\
have been conducted using O1 and O2 data
(\citealt{fesikpapa20,fesikpapa20b}; see also \citealt{carideetal19}).
We estimate, using the GW $h_0$ sensitivity upper limit of O2 data
at $\nugw\approx90\mbox{ Hz}$ from
\citet{abbottetal19c}, that an improved targeted search of r-modes
in \psr\ using a contemporaneous timing model and accounting for
glitches would yield $h_0/\hsd<0.3$,
which would set the limit $\alpha<0.06$
and limit the percentage of rotational energy loss that is due to
GW emission to $<10$~percent.
Efforts are underway to search the latest, most sensitive GW data
from the third observing run (O3), which collected data from
2019 April 1 to 2020 March 27.
The timing model provided here using \NICER\ data, with 3 glitches
occurring during O3 (see Figure~\ref{fig:glitch}), will produce the
best limits on ellipticity and r-mode amplitude of \psr.

\section*{Acknowledgements}

The authors thank the anonymous referee for comments which led to
improvements in the manuscript.
WCGH greatly appreciates advice and support on timing analyses by
S. Bogdanov, A. Lommen, and E. Teng
and thanks N. Andersson and D.I. Jones for discussions and
A. Beri, A.K. Harding, A. Kr\'olak, S. Mastrogiovanni, M.A. Papa,
M. Pitkin, G. Raman, K. Riles, and G. Woan for support.
WCGH acknowledges support through grant 80NSSC19K1444 from NASA.
CME acknowledges support from FONDECYT/Regular 1171421 and
USA1899-Vridei 041931SSSA-PAP (Universidad de Santiago de Chile, USACH).
MB is partially supported by Polish NCN grant no.~2017/26/M/ST9/00978.
BH is supported by Polish NCN grant no.~2015/18/E/ST9/00577.
This work is supported by NASA through the \NICER\ mission and the
Astrophysics Explorers Program and makes use of data and software
provided by the High Energy Astrophysics Science Archive Research Center
(HEASARC), which is a service of the Astrophysics Science Division at
NASA/GSFC and High Energy Astrophysics Division of the Smithsonian
Astrophysical Observatory.


\section*{Data availability}

The data underlying this article will be shared on reasonable
request to the corresponding author.


\bibliographystyle{mnras}
\bibliography{psrj0537}

\begin{thebibliography}{}
\makeatletter
\relax
\def\mn@urlcharsother{\let\do\@makeother \do\$\do\&\do\#\do\^\do\_\do\%\do\~}
\def\mn@doi{\begingroup\mn@urlcharsother \@ifnextchar [ {\mn@doi@}
  {\mn@doi@[]}}
\def\mn@doi@[#1]#2{\def\@tempa{#1}\ifx\@tempa\@empty \href
  {http://dx.doi.org/#2} {doi:#2}\else \href {http://dx.doi.org/#2} {#1}\fi
  \endgroup}
\def\mn@eprint#1#2{\mn@eprint@#1:#2::\@nil}
\def\mn@eprint@arXiv#1{\href {http://arxiv.org/abs/#1} {{\tt arXiv:#1}}}
\def\mn@eprint@dblp#1{\href {http://dblp.uni-trier.de/rec/bibtex/#1.xml}
  {dblp:#1}}
\def\mn@eprint@#1:#2:#3:#4\@nil{\def\@tempa {#1}\def\@tempb {#2}\def\@tempc
  {#3}\ifx \@tempc \@empty \let \@tempc \@tempb \let \@tempb \@tempa \fi \ifx
  \@tempb \@empty \def\@tempb {arXiv}\fi \@ifundefined
  {mn@eprint@\@tempb}{\@tempb:\@tempc}{\expandafter \expandafter \csname
  mn@eprint@\@tempb\endcsname \expandafter{\@tempc}}}

\bibitem[\protect\citeauthoryear{{{Abbott} et al.}}{{{Abbott} et
  al.}}{2019a}]{abbottetal19a}
{{Abbott} et al.} 2019a, \mn@doi [\prd] {10.1103/PhysRevD.99.122002}, \href
  {https://ui.adsabs.harvard.edu/abs/2019PhRvD..99l2002A} {99, 122002}

\bibitem[\protect\citeauthoryear{{{Abbott} et al.}}{{{Abbott} et
  al.}}{2019b}]{abbottetal19b}
{{Abbott} et al.} 2019b, \mn@doi [\apj] {10.3847/1538-4357/ab113b}, \href
  {http://adsabs.harvard.edu/abs/2019ApJ...875..122A} {875, 122}

\bibitem[\protect\citeauthoryear{{{Abbott} et al.}}{{{Abbott} et
  al.}}{2019c}]{abbottetal19c}
{{Abbott} et al.} 2019c, \mn@doi [\apj] {10.3847/1538-4357/ab20cb}, \href
  {https://ui.adsabs.harvard.edu/abs/2019ApJ...879...10A} {879, 10}

\bibitem[\protect\citeauthoryear{{Akbal}, {Alpar}, {Buchner}  \&
  {Pines}}{{Akbal} et~al.}{2017}]{akbaletal17}
{Akbal} O.,  {Alpar} M.~A.,  {Buchner} S.,   {Pines} D.,  2017, \mn@doi
  [\mnras] {10.1093/mnras/stx1095}, \href
  {https://ui.adsabs.harvard.edu/abs/2017MNRAS.469.4183A} {469, 4183}

\bibitem[\protect\citeauthoryear{{Alford} \& {Schwenzer}}{{Alford} \&
  {Schwenzer}}{2014}]{alfordschwenzer14}
{Alford} M.~G.,  {Schwenzer} K.,  2014, \mn@doi [\apj]
  {10.1088/0004-637X/781/1/26}, \href
  {https://ui.adsabs.harvard.edu/abs/2014ApJ...781...26A} {781, 26}

\bibitem[\protect\citeauthoryear{{Alpar}, {Pines}, {Anderson}  \&
  {Shaham}}{{Alpar} et~al.}{1984}]{alparetal84}
{Alpar} M.~A.,  {Pines} D.,  {Anderson} P.~W.,   {Shaham} J.,  1984, \mn@doi
  [\apj] {10.1086/161616}, \href
  {https://ui.adsabs.harvard.edu/abs/1984ApJ...276..325A} {276, 325}

\bibitem[\protect\citeauthoryear{{Anderson} \& {Itoh}}{{Anderson} \&
  {Itoh}}{1975}]{andersonitoh75}
{Anderson} P.~W.,  {Itoh} N.,  1975, \mn@doi [\nat] {10.1038/256025a0}, \href
  {https://ui.adsabs.harvard.edu/abs/1975Natur.256...25A} {256, 25}

\bibitem[\protect\citeauthoryear{{Andersson}}{{Andersson}}{1998}]{andersson98}
{Andersson} N.,  1998, \mn@doi [\apj] {10.1086/305919}, \href
  {https://ui.adsabs.harvard.edu/abs/1998ApJ...502..708A} {502, 708}

\bibitem[\protect\citeauthoryear{{Andersson} \& {Kokkotas}}{{Andersson} \&
  {Kokkotas}}{2001}]{anderssonkokkotas01}
{Andersson} N.,  {Kokkotas} K.~D.,  2001, \mn@doi [International Journal of
  Modern Physics D] {10.1142/S0218271801001062}, \href
  {https://ui.adsabs.harvard.edu/abs/2001IJMPD..10..381A} {10, 381}

\bibitem[\protect\citeauthoryear{{Andersson}, {Glampedakis}, {Ho}  \&
  {Espinoza}}{{Andersson} et~al.}{2012}]{anderssonetal12}
{Andersson} N.,  {Glampedakis} K.,  {Ho} W.~C.~G.,   {Espinoza} C.~M.,  2012,
  \mn@doi [\prl] {10.1103/PhysRevLett.109.241103}, \href
  {https://ui.adsabs.harvard.edu/abs/2012PhRvL.109x1103A} {109, 241103}

\bibitem[\protect\citeauthoryear{{Andersson}, {Jones}  \& {Ho}}{{Andersson}
  et~al.}{2014}]{anderssonetal14}
{Andersson} N.,  {Jones} D.~I.,   {Ho} W.~C.~G.,  2014, \mn@doi [\mnras]
  {10.1093/mnras/stu870}, \href
  {https://ui.adsabs.harvard.edu/abs/2014MNRAS.442.1786A} {442, 1786}

\bibitem[\protect\citeauthoryear{{Andersson}, {Antonopoulou}, {Espinoza},
  {Haskell}  \& {Ho}}{{Andersson} et~al.}{2018}]{anderssonetal18}
{Andersson} N.,  {Antonopoulou} D.,  {Espinoza} C.~M.,  {Haskell} B.,   {Ho}
  W.~C.~G.,  2018, \mn@doi [\apj] {10.3847/1538-4357/aad6eb}, \href
  {https://ui.adsabs.harvard.edu/abs/2018ApJ...864..137A} {864, 137}

\bibitem[\protect\citeauthoryear{{Antonopoulou}, {Espinoza}, {Kuiper}  \&
  {Andersson}}{{Antonopoulou} et~al.}{2018}]{antonopoulouetal18}
{Antonopoulou} D.,  {Espinoza} C.~M.,  {Kuiper} L.,   {Andersson} N.,  2018,
  \mn@doi [\mnras] {10.1093/mnras/stx2429}, \href
  {https://ui.adsabs.harvard.edu/abs/2018MNRAS.473.1644A} {473, 1644}

\bibitem[\protect\citeauthoryear{{Ashton}, {Prix}  \& {Jones}}{{Ashton}
  et~al.}{2017}]{ashtonetal17}
{Ashton} G.,  {Prix} R.,   {Jones} D.~I.,  2017, \mn@doi [\prd]
  {10.1103/PhysRevD.96.063004}, \href
  {https://ui.adsabs.harvard.edu/abs/2017PhRvD..96f3004A} {96, 063004}

\bibitem[\protect\citeauthoryear{{Caride}, {Inta}, {Owen}  \& {Rajbhand
  ari}}{{Caride} et~al.}{2019}]{carideetal19}
{Caride} S.,  {Inta} R.,  {Owen} B.~J.,   {Rajbhand ari} B.,  2019, \mn@doi
  [\prd] {10.1103/PhysRevD.100.064013}, \href
  {https://ui.adsabs.harvard.edu/abs/2019PhRvD.100f4013C} {100, 064013}

\bibitem[\protect\citeauthoryear{{Chamel}}{{Chamel}}{2012}]{chamel12}
{Chamel} N.,  2012, \mn@doi [\prc] {10.1103/PhysRevC.85.035801}, \href
  {https://ui.adsabs.harvard.edu/abs/2012PhRvC..85c5801C} {85, 035801}

\bibitem[\protect\citeauthoryear{{Chamel}}{{Chamel}}{2013}]{chamel13}
{Chamel} N.,  2013, \mn@doi [\prl] {10.1103/PhysRevLett.110.011101}, \href
  {https://ui.adsabs.harvard.edu/abs/2013PhRvL.110a1101C} {110, 011101}

\bibitem[\protect\citeauthoryear{{Chamel}}{{Chamel}}{2017}]{chamel17}
{Chamel} N.,  2017, \mn@doi [Journal of Low Temperature Physics]
  {10.1007/s10909-017-1815-x}, \href
  {https://ui.adsabs.harvard.edu/abs/2017JLTP..189..328C} {189, 328}

\bibitem[\protect\citeauthoryear{{Chen}, {Wang}, {Gotthelf}, {Jiang}, {Chu}  \&
  {Gruendl}}{{Chen} et~al.}{2006}]{chenetal06}
{Chen} Y.,  {Wang} Q.~D.,  {Gotthelf} E.~V.,  {Jiang} B.,  {Chu} Y.-H.,
  {Gruendl} R.,  2006, \mn@doi [\apj] {10.1086/507017}, \href
  {https://ui.adsabs.harvard.edu/abs/2006ApJ...651..237C} {651, 237}

\bibitem[\protect\citeauthoryear{{Dodson}, {Lewis}  \& {McCulloch}}{{Dodson}
  et~al.}{2007}]{dodsonetal07}
{Dodson} R.,  {Lewis} D.,   {McCulloch} P.,  2007, \mn@doi [\apss]
  {10.1007/s10509-007-9372-4}, \href
  {https://ui.adsabs.harvard.edu/abs/2007Ap&SS.308..585D} {308, 585}

\bibitem[\protect\citeauthoryear{{Edwards}, {Hobbs}  \& {Manchester}}{{Edwards}
  et~al.}{2006}]{edwardsetal06}
{Edwards} R.~T.,  {Hobbs} G.~B.,   {Manchester} R.~N.,  2006, \mn@doi [\mnras]
  {10.1111/j.1365-2966.2006.10870.x}, \href
  {https://ui.adsabs.harvard.edu/abs/2006MNRAS.372.1549E} {372, 1549}

\bibitem[\protect\citeauthoryear{{Espinoza}, {Lyne}, {Stappers}  \&
  {Kramer}}{{Espinoza} et~al.}{2011}]{espinozaetal11}
{Espinoza} C.~M.,  {Lyne} A.~G.,  {Stappers} B.~W.,   {Kramer} M.,  2011,
  \mn@doi [\mnras] {10.1111/j.1365-2966.2011.18503.x}, \href
  {https://ui.adsabs.harvard.edu/abs/2011MNRAS.414.1679E} {414, 1679}

\bibitem[\protect\citeauthoryear{{Espinoza}, {Lyne}  \& {Stappers}}{{Espinoza}
  et~al.}{2017}]{espinozaetal17}
{Espinoza} C.~M.,  {Lyne} A.~G.,   {Stappers} B.~W.,  2017, \mn@doi [\mnras]
  {10.1093/mnras/stw3081}, \href
  {https://ui.adsabs.harvard.edu/abs/2017MNRAS.466..147E} {466, 147}

\bibitem[\protect\citeauthoryear{{Feng} et~al.,}{{Feng}
  et~al.}{2020}]{fengetal20}
{Feng} H.,  et~al., 2020, \mn@doi [Nature Astronomy]
  {10.1038/s41550-020-1088-1}, \href
  {https://ui.adsabs.harvard.edu/abs/2020NatAs...4..511F} {4, 511}

\bibitem[\protect\citeauthoryear{{Ferdman}, {Archibald}, {Gourgouliatos}  \&
  {Kaspi}}{{Ferdman} et~al.}{2018}]{ferdmanetal18}
{Ferdman} R.~D.,  {Archibald} R.~F.,  {Gourgouliatos} K.~N.,   {Kaspi} V.~M.,
  2018, \mn@doi [\apj] {10.3847/1538-4357/aaa198}, \href
  {https://ui.adsabs.harvard.edu/abs/2018ApJ...852..123F} {852, 123}

\bibitem[\protect\citeauthoryear{{Fesik} \& {Papa}}{{Fesik} \&
  {Papa}}{2020a}]{fesikpapa20}
{Fesik} L.,  {Papa} M.~A.,  2020a, \mn@doi [\apj] {10.3847/1538-4357/ab8193},
  \href {https://ui.adsabs.harvard.edu/abs/2020ApJ...895...11F} {895, 11}

\bibitem[\protect\citeauthoryear{{Fesik} \& {Papa}}{{Fesik} \&
  {Papa}}{2020b}]{fesikpapa20b}
{Fesik} L.,  {Papa} M.~A.,  2020b, \mn@doi [\apj] {10.3847/1538-4357/aba04e},
  \href {https://ui.adsabs.harvard.edu/abs/2020ApJ...897..185F} {897, 185}

\bibitem[\protect\citeauthoryear{{Friedman} \& {Morsink}}{{Friedman} \&
  {Morsink}}{1998}]{friedmanmorsink98}
{Friedman} J.~L.,  {Morsink} S.~M.,  1998, \mn@doi [\apj] {10.1086/305920},
  \href {https://ui.adsabs.harvard.edu/abs/1998ApJ...502..714F} {502, 714}

\bibitem[\protect\citeauthoryear{{Fuentes}, {Espinoza}, {Reisenegger}, {Shaw},
  {Stappers}  \& {Lyne}}{{Fuentes} et~al.}{2017}]{fuentesetal17}
{Fuentes} J.~R.,  {Espinoza} C.~M.,  {Reisenegger} A.,  {Shaw} B.,  {Stappers}
  B.~W.,   {Lyne} A.~G.,  2017, \mn@doi [\aap] {10.1051/0004-6361/201731519},
  \href {https://ui.adsabs.harvard.edu/abs/2017A&A...608A.131F} {608, A131}

\bibitem[\protect\citeauthoryear{{Fuentes}, {Espinoza}  \&
  {Reisenegger}}{{Fuentes} et~al.}{2019}]{fuentesetal19}
{Fuentes} J.~R.,  {Espinoza} C.~M.,   {Reisenegger} A.,  2019, \mn@doi [\aap]
  {10.1051/0004-6361/201935939}, \href
  {https://ui.adsabs.harvard.edu/abs/2019A&A...630A.115F} {630, A115}

\bibitem[\protect\citeauthoryear{{Gourgouliatos} \& {Cumming}}{{Gourgouliatos}
  \& {Cumming}}{2015}]{gourgouliatoscumming15}
{Gourgouliatos} K.~N.,  {Cumming} A.,  2015, \mn@doi [\mnras]
  {10.1093/mnras/stu2140}, \href
  {https://ui.adsabs.harvard.edu/abs/2015MNRAS.446.1121G} {446, 1121}

\bibitem[\protect\citeauthoryear{{Graber}, {Andersson}  \& {Hogg}}{{Graber}
  et~al.}{2017}]{graberetal17}
{Graber} V.,  {Andersson} N.,   {Hogg} M.,  2017, \mn@doi [International
  Journal of Modern Physics D] {10.1142/S0218271817300154}, \href
  {https://ui.adsabs.harvard.edu/abs/2017IJMPD..2630015G} {26, 1730015}

\bibitem[\protect\citeauthoryear{{Haskell} \& {Melatos}}{{Haskell} \&
  {Melatos}}{2015}]{haskellmelatos15}
{Haskell} B.,  {Melatos} A.,  2015, \mn@doi [International Journal of Modern
  Physics D] {10.1142/S0218271815300086}, \href
  {https://ui.adsabs.harvard.edu/abs/2015IJMPD..2430008H} {24, 1530008}

\bibitem[\protect\citeauthoryear{{Ho}}{{Ho}}{2015}]{ho15}
{Ho} W. C.~G.,  2015, \mn@doi [\mnras] {10.1093/mnras/stv1339}, \href
  {https://ui.adsabs.harvard.edu/abs/2015MNRAS.452..845H} {452, 845}

\bibitem[\protect\citeauthoryear{{Ho} \& {Andersson}}{{Ho} \&
  {Andersson}}{2012}]{hoandersson12}
{Ho} W. C.~G.,  {Andersson} N.,  2012, \mn@doi [Nature Physics]
  {10.1038/nphys2424}, \href
  {https://ui.adsabs.harvard.edu/abs/2012NatPh...8..787H} {8, 787}

\bibitem[\protect\citeauthoryear{{Ho}, {Espinoza}, {Antonopoulou}  \&
  {Andersson}}{{Ho} et~al.}{2015}]{hoetal15}
{Ho} W.~C.~G.,  {Espinoza} C.~M.,  {Antonopoulou} D.,   {Andersson} N.,  2015,
  \mn@doi [Science Advances] {10.1126/sciadv.1500578}, \href
  {https://ui.adsabs.harvard.edu/abs/2015SciA....1E0578H} {1, e1500578}

\bibitem[\protect\citeauthoryear{{Ho}, {Espinoza}, {Antonopoulou}  \&
  {Andersson}}{{Ho} et~al.}{2017}]{hoetal17}
{Ho} W. C.~G.,  {Espinoza} C.~M.,  {Antonopoulou} D.,   {Andersson} N.,  2017,
  in {Kubono} S.,  {Kajino} T.,  {Nishimura} S.,  {Isobe} T.,  {Nagataki} S.,
  {Shima} T.,   {Takeda} Y.,  eds, 14th International Symposium on Nuclei in
  the Cosmos (NIC2016). p. 010805 (\mn@eprint {arXiv} {1703.00932}),
  \mn@doi{10.7566/JPSCP.14.010805}

\bibitem[\protect\citeauthoryear{{Ho}, {Heinke}  \& {Chugunov}}{{Ho}
  et~al.}{2019}]{hoetal19}
{Ho} W. C.~G.,  {Heinke} C.~O.,   {Chugunov} A.~I.,  2019, \mn@doi [\apj]
  {10.3847/1538-4357/ab3578}, \href
  {https://ui.adsabs.harvard.edu/abs/2019ApJ...882..128H} {882, 128}

\bibitem[\protect\citeauthoryear{{Ho}, {Jones}, {Andersson}  \&
  {Espinoza}}{{Ho} et~al.}{2020}]{hoetal20}
{Ho} W. C.~G.,  {Jones} D.~I.,  {Andersson} N.,   {Espinoza} C.~M.,  2020,
  \mn@doi [\prd] {10.1103/PhysRevD.101.103009}, \href
  {https://ui.adsabs.harvard.edu/abs/2020PhRvD.101j3009H} {101, 103009}

\bibitem[\protect\citeauthoryear{{Hobbs}, {Edwards}  \& {Manchester}}{{Hobbs}
  et~al.}{2006}]{hobbsetal06}
{Hobbs} G.~B.,  {Edwards} R.~T.,   {Manchester} R.~N.,  2006, \mn@doi [\mnras]
  {10.1111/j.1365-2966.2006.10302.x}, \href
  {https://ui.adsabs.harvard.edu/abs/2006MNRAS.369..655H} {369, 655}

\bibitem[\protect\citeauthoryear{{Howitt}, {Melatos}  \& {Delaigle}}{{Howitt}
  et~al.}{2018}]{howittetal18}
{Howitt} G.,  {Melatos} A.,   {Delaigle} A.,  2018, \mn@doi [\apj]
  {10.3847/1538-4357/aae20a}, \href
  {https://ui.adsabs.harvard.edu/abs/2018ApJ...867...60H} {867, 60}

\bibitem[\protect\citeauthoryear{{Idrisy}, {Owen}  \& {Jones}}{{Idrisy}
  et~al.}{2015}]{idrisyetal15}
{Idrisy} A.,  {Owen} B.~J.,   {Jones} D.~I.,  2015, \mn@doi [\prd]
  {10.1103/PhysRevD.91.024001}, \href
  {https://ui.adsabs.harvard.edu/abs/2015PhRvD..91b4001I} {91, 024001}

\bibitem[\protect\citeauthoryear{{Jasiulek} \& {Chirenti}}{{Jasiulek} \&
  {Chirenti}}{2017}]{jasiulekchirenti17}
{Jasiulek} M.,  {Chirenti} C.,  2017, \mn@doi [\prd]
  {10.1103/PhysRevD.95.064060}, \href
  {https://ui.adsabs.harvard.edu/abs/2017PhRvD..95f4060J} {95, 064060}

\bibitem[\protect\citeauthoryear{{Johnston} \& {Karastergiou}}{{Johnston} \&
  {Karastergiou}}{2017}]{johnstonkarastergiou17}
{Johnston} S.,  {Karastergiou} A.,  2017, \mn@doi [\mnras]
  {10.1093/mnras/stx377}, \href
  {https://ui.adsabs.harvard.edu/abs/2017MNRAS.467.3493J} {467, 3493}

\bibitem[\protect\citeauthoryear{{Keitel} et~al.,}{{Keitel}
  et~al.}{2019}]{keiteletal19}
{Keitel} D.,  et~al., 2019, \mn@doi [\prd] {10.1103/PhysRevD.100.064058}, \href
  {https://ui.adsabs.harvard.edu/abs/2019PhRvD.100f4058K} {100, 064058}

\bibitem[\protect\citeauthoryear{{Kuiper} \& {Hermsen}}{{Kuiper} \&
  {Hermsen}}{2015}]{kuiperhermsen15}
{Kuiper} L.,  {Hermsen} W.,  2015, \mn@doi [\mnras] {10.1093/mnras/stv426},
  \href {https://ui.adsabs.harvard.edu/abs/2015MNRAS.449.3827K} {449, 3827}

\bibitem[\protect\citeauthoryear{{Link}, {Epstein}  \& {Lattimer}}{{Link}
  et~al.}{1999}]{linketal99}
{Link} B.,  {Epstein} R.~I.,   {Lattimer} J.~M.,  1999, \mn@doi [\prl]
  {10.1103/PhysRevLett.83.3362}, \href
  {https://ui.adsabs.harvard.edu/abs/1999PhRvL..83.3362L} {83, 3362}

\bibitem[\protect\citeauthoryear{{Lyne}, {Jordan}, {Graham-Smith}, {Espinoza},
  {Stappers}  \& {Weltevrede}}{{Lyne} et~al.}{2015}]{lyneetal15}
{Lyne} A.~G.,  {Jordan} C.~A.,  {Graham-Smith} F.,  {Espinoza} C.~M.,
  {Stappers} B.~W.,   {Weltevrede} P.,  2015, \mn@doi [\mnras]
  {10.1093/mnras/stu2118}, \href
  {https://ui.adsabs.harvard.edu/abs/2015MNRAS.446..857L} {446, 857}

\bibitem[\protect\citeauthoryear{{Manchester}, {Hobbs}, {Teoh}  \&
  {Hobbs}}{{Manchester} et~al.}{2005}]{manchesteretal05}
{Manchester} R.~N.,  {Hobbs} G.~B.,  {Teoh} A.,   {Hobbs} M.,  2005, \mn@doi
  [\aj] {10.1086/428488}, \href
  {https://ui.adsabs.harvard.edu/abs/2005AJ....129.1993M} {129, 1993}

\bibitem[\protect\citeauthoryear{{Marshall}, {Gotthelf}, {Zhang}, {Middleditch}
   \& {Wang}}{{Marshall} et~al.}{1998}]{marshalletal98}
{Marshall} F.~E.,  {Gotthelf} E.~V.,  {Zhang} W.,  {Middleditch} J.,   {Wang}
  Q.~D.,  1998, \mn@doi [\apjl] {10.1086/311381}, \href
  {https://ui.adsabs.harvard.edu/abs/1998ApJ...499L.179M} {499, L179}

\bibitem[\protect\citeauthoryear{{Marshall}, {Gotthelf}, {Middleditch}, {Wang}
  \& {Zhang}}{{Marshall} et~al.}{2004}]{marshalletal04}
{Marshall} F.~E.,  {Gotthelf} E.~V.,  {Middleditch} J.,  {Wang} Q.~D.,
  {Zhang} W.,  2004, \mn@doi [\apj] {10.1086/381567}, \href
  {https://ui.adsabs.harvard.edu/abs/2004ApJ...603..682M} {603, 682}

\bibitem[\protect\citeauthoryear{{McKenna} \& {Lyne}}{{McKenna} \&
  {Lyne}}{1990}]{mckennalyne90}
{McKenna} J.,  {Lyne} A.~G.,  1990, \mn@doi [\nat] {10.1038/343349a0}, \href
  {https://ui.adsabs.harvard.edu/abs/1990Natur.343..349M} {343, 349}

\bibitem[\protect\citeauthoryear{{Melatos} \& {Drummond}}{{Melatos} \&
  {Drummond}}{2019}]{melatosdrummond19}
{Melatos} A.,  {Drummond} L.~V.,  2019, \mn@doi [\apj]
  {10.3847/1538-4357/ab44c3}, \href
  {https://ui.adsabs.harvard.edu/abs/2019ApJ...885...37M} {885, 37}

\bibitem[\protect\citeauthoryear{{Melatos}, {Peralta}  \& {Wyithe}}{{Melatos}
  et~al.}{2008}]{melatosetal08}
{Melatos} A.,  {Peralta} C.,   {Wyithe} J.~S.~B.,  2008, \mn@doi [\apj]
  {10.1086/523349}, \href
  {https://ui.adsabs.harvard.edu/abs/2008ApJ...672.1103M} {672, 1103}

\bibitem[\protect\citeauthoryear{{Melatos}, {Howitt}  \& {Fulgenzi}}{{Melatos}
  et~al.}{2018}]{melatosetal18}
{Melatos} A.,  {Howitt} G.,   {Fulgenzi} W.,  2018, \mn@doi [\apj]
  {10.3847/1538-4357/aad228}, \href
  {https://ui.adsabs.harvard.edu/abs/2018ApJ...863..196M} {863, 196}

\bibitem[\protect\citeauthoryear{{Michel}}{{Michel}}{1969}]{michel69}
{Michel} F.~C.,  1969, \mn@doi [\apj] {10.1086/150233}, \href
  {https://ui.adsabs.harvard.edu/abs/1969ApJ...158..727M} {158, 727}

\bibitem[\protect\citeauthoryear{{Michel} \& {Tucker}}{{Michel} \&
  {Tucker}}{1969}]{micheltucker69}
{Michel} F.~C.,  {Tucker} W.~H.,  1969, \mn@doi [\nat] {10.1038/223277a0},
  \href {https://ui.adsabs.harvard.edu/abs/1969Natur.223..277M} {223, 277}

\bibitem[\protect\citeauthoryear{{Middleditch}, {Marshall}, {Wang}, {Gotthelf}
  \& {Zhang}}{{Middleditch} et~al.}{2006}]{middleditchetal06}
{Middleditch} J.,  {Marshall} F.~E.,  {Wang} Q.~D.,  {Gotthelf} E.~V.,
  {Zhang} W.,  2006, \mn@doi [\apj] {10.1086/508736}, \href
  {https://ui.adsabs.harvard.edu/abs/2006ApJ...652.1531M} {652, 1531}

\bibitem[\protect\citeauthoryear{{Owen}}{{Owen}}{2010}]{owen10}
{Owen} B.~J.,  2010, \mn@doi [\prd] {10.1103/PhysRevD.82.104002}, \href
  {https://ui.adsabs.harvard.edu/abs/2010PhRvD..82j4002O} {82, 104002}

\bibitem[\protect\citeauthoryear{{Owen}, {Lindblom}, {Cutler}, {Schutz},
  {Vecchio}  \& {Andersson}}{{Owen} et~al.}{1998}]{owenetal98}
{Owen} B.~J.,  {Lindblom} L.,  {Cutler} C.,  {Schutz} B.~F.,  {Vecchio} A.,
  {Andersson} N.,  1998, \mn@doi [\prd] {10.1103/PhysRevD.58.084020}, \href
  {http://adsabs.harvard.edu/abs/1998PhRvD..58h4020O} {58, 084020}

\bibitem[\protect\citeauthoryear{{Palfreyman}, {Dickey}, {Hotan}, {Ellingsen}
  \& {van Straten}}{{Palfreyman} et~al.}{2018}]{palfreymanetal18}
{Palfreyman} J.,  {Dickey} J.~M.,  {Hotan} A.,  {Ellingsen} S.,   {van Straten}
  W.,  2018, \mn@doi [\nat] {10.1038/s41586-018-0001-x}, \href
  {https://ui.adsabs.harvard.edu/abs/2018Natur.556..219P} {556, 219}

\bibitem[\protect\citeauthoryear{{Parthasarathy} et~al.,}{{Parthasarathy}
  et~al.}{2020}]{parthasarathyetal20}
{Parthasarathy} A.,  et~al., 2020, \mn@doi [\mnras] {10.1093/mnras/staa882},
  \href {https://ui.adsabs.harvard.edu/abs/2020MNRAS.494.2012P} {494, 2012}

\bibitem[\protect\citeauthoryear{{Pietrzy{\'n}ski} et~al.,}{{Pietrzy{\'n}ski}
  et~al.}{2019}]{pietrzynskietal19}
{Pietrzy{\'n}ski} G.,  et~al., 2019, \mn@doi [\nat]
  {10.1038/s41586-019-0999-4}, \href
  {https://ui.adsabs.harvard.edu/abs/2019Natur.567..200P} {567, 200}

\bibitem[\protect\citeauthoryear{{Ransom}, {Eikenberry}  \&
  {Middleditch}}{{Ransom} et~al.}{2002}]{ransometal02}
{Ransom} S.~M.,  {Eikenberry} S.~S.,   {Middleditch} J.,  2002, \mn@doi [\aj]
  {10.1086/342285}, \href
  {https://ui.adsabs.harvard.edu/abs/2002AJ....124.1788R} {124, 1788}

\bibitem[\protect\citeauthoryear{{Ray} et~al.,}{{Ray} et~al.}{2019}]{rayetal19}
{Ray} P.~S.,  et~al., 2019, \mn@doi [\apj] {10.3847/1538-4357/ab24d8}, \href
  {https://ui.adsabs.harvard.edu/abs/2019ApJ...879..130R} {879, 130}

\bibitem[\protect\citeauthoryear{{Romani}}{{Romani}}{1990}]{romani90}
{Romani} R.~W.,  1990, \mn@doi [\nat] {10.1038/347741a0}, \href
  {https://ui.adsabs.harvard.edu/abs/1990Natur.347..741R} {347, 741}

\bibitem[\protect\citeauthoryear{{Sauls}, {Chamel}  \& {Alpar}}{{Sauls}
  et~al.}{2020}]{saulsetal20}
{Sauls} J.~A.,  {Chamel} N.,   {Alpar} M.~A.,  2020, arXiv e-prints, \href
  {https://ui.adsabs.harvard.edu/abs/2020arXiv200109959S} {p. arXiv:2001.09959}

\bibitem[\protect\citeauthoryear{{Shapiro} \& {Teukolsky}}{{Shapiro} \&
  {Teukolsky}}{1983}]{shapiroteukolsky83}
{Shapiro} S.~L.,  {Teukolsky} S.~A.,  1983, {Black holes, white dwarfs, and
  neutron stars : the physics of compact objects}.
Wiley

\bibitem[\protect\citeauthoryear{{Tong}, {Xu}, {Song}  \& {Qiao}}{{Tong}
  et~al.}{2013}]{tongetal13}
{Tong} H.,  {Xu} R.~X.,  {Song} L.~M.,   {Qiao} G.~J.,  2013, \mn@doi [\apj]
  {10.1088/0004-637X/768/2/144}, \href
  {https://ui.adsabs.harvard.edu/abs/2013ApJ...768..144T} {768, 144}

\bibitem[\protect\citeauthoryear{{Townsley}, {Broos}, {Feigelson}, {Garmire}
  \& {Getman}}{{Townsley} et~al.}{2006}]{townsleyetal06}
{Townsley} L.~K.,  {Broos} P.~S.,  {Feigelson} E.~D.,  {Garmire} G.~P.,
  {Getman} K.~V.,  2006, \mn@doi [\aj] {10.1086/500535}, \href
  {https://ui.adsabs.harvard.edu/abs/2006AJ....131.2164T} {131, 2164}

\bibitem[\protect\citeauthoryear{{Wang} \& {Gotthelf}}{{Wang} \&
  {Gotthelf}}{1998}]{wangetal98}
{Wang} Q.~D.,  {Gotthelf} E.~V.,  1998, \mn@doi [\apj] {10.1086/305214}, \href
  {https://ui.adsabs.harvard.edu/abs/1998ApJ...494..623W} {494, 623}

\bibitem[\protect\citeauthoryear{{Yu} et~al.,}{{Yu} et~al.}{2013}]{yuetal13}
{Yu} M.,  et~al., 2013, \mn@doi [\mnras] {10.1093/mnras/sts366}, \href
  {https://ui.adsabs.harvard.edu/abs/2013MNRAS.429..688Y} {429, 688}

\makeatother
\end{thebibliography}


\bsp	
\label{lastpage}
\end{document}